\documentclass[10pt,draftcls,peerreview]{IEEEtran}
\usepackage{cite,graphicx,psfrag,amsmath,amssymb,url}
\newtheorem{theorem}{Theorem}
\newtheorem{lemma}{Lemma}

\newtheorem{definition}{Definition}

\def\algd{2}
\def\infd{D}
\psfrag{-1}{$-1$}
\psfrag{0}{$0$}
\psfrag{pval1}{$p = (\frac{4}{9},\frac{1}{4},\frac{19}{180},\frac{1}{10},\frac{1}{10})$}
\psfrag{pval2}{$p = (0.58,0.12,0.11,0.1,0.09)$}
\psfrag{4o9}{$\frac{4}{9}$}
\psfrag{1o4}{$\frac{1}{4}$}
\psfrag{19o180}{$\frac{19}{180}$}
\psfrag{1o10}{$\frac{1}{10}$}
\psfrag{2neg1}{$\algd^{-1}$}
\psfrag{2negmid}{$\algd^{-l_\midt}$}
\psfrag{2negL}{$\algd^{-l_{\max}}$}
\psfrag{lmid}{$l_\mid$}
\psfrag{n}{$n$}
\psfrag{n-m}{$n-m$}
\psfrag{L}{$l_{\max}$}
\psfrag{1}{$1$}
\psfrag{infty}{$\infty$}
\psfrag{1}{$1$}
\psfrag{2}{$2$}
\psfrag{3}{$3$}
\psfrag{00}{$00$}
\psfrag{01}{$01$}
\psfrag{10}{$10$}
\psfrag{11}{$11$}
\psfrag{001}{$001$}
\psfrag{000}{$000$}
\psfrag{l (level)}{$l$ (level)}
\psfrag{i (item)}{$i$ (item)}
\psfrag{(width)}{$\rho$ (width)}
\psfrag{1m1}{\scriptsize $\mu(1,1) = p_1$}
\psfrag{2m1}{\scriptsize $\mu(2,1) = p_2$}
\psfrag{3m1}{\scriptsize $\mu(3,1) = p_3$}
\psfrag{4m1}{\scriptsize $\mu(4,1) = p_4$}
\psfrag{1m2}{\scriptsize $\mu(1,2) = 3p_1$}
\psfrag{2m2}{\scriptsize $\mu(2,2) = 3p_2$}
\psfrag{3m2}{\scriptsize $\mu(3,2) = 3p_3$}
\psfrag{4m2}{\scriptsize $\mu(4,2) = 3p_4$}
\psfrag{1m3}{\scriptsize $\mu(1,3) = 5p_1$}
\psfrag{2m3}{\scriptsize $\mu(2,3) = 5p_2$}
\psfrag{3m3}{\scriptsize $\mu(3,3) = 5p_3$}
\psfrag{4m3}{\scriptsize $\mu(4,3) = 5p_4$}
\psfrag{1r1}{\scriptsize $\rho(1,1) = \frac{1}{2}$}
\psfrag{1r2}{\scriptsize $\rho(2,1) = \frac{1}{2}$}
\psfrag{1r3}{\scriptsize $\rho(3,1) = \frac{1}{2}$}
\psfrag{1r4}{\scriptsize $\rho(4,1) = \frac{1}{2}$}
\psfrag{2r1}{\scriptsize $\rho(1,2) = \frac{1}{4}$}
\psfrag{2r2}{\scriptsize $\rho(2,2) = \frac{1}{4}$}
\psfrag{2r3}{\scriptsize $\rho(3,2) = \frac{1}{4}$}
\psfrag{2r4}{\scriptsize $\rho(4,2) = \frac{1}{4}$}
\psfrag{3r1}{\scriptsize $\rho(1,3) = \frac{1}{8}$}
\psfrag{3r2}{\scriptsize $\rho(2,3) = \frac{1}{8}$}
\psfrag{3r3}{\scriptsize $\rho(3,3) = \frac{1}{8}$}
\psfrag{3r4}{\scriptsize $\rho(4,3) = \frac{1}{8}$}
\psfrag{=0=0}{\large $t=0=0_2$}
\psfrag{=2=10}{\large $t=2=10_2$}
\psfrag{=3=11}{\large $t=3=11_2$}
\psfrag{m=1}{\large $\mu=1$}
\psfrag{m=2}{\large $\mu=2$}
\psfrag{m=4}{\large $\mu=4$}
\psfrag{m=5}{\large $\mu=5$}
\psfrag{m=6}{\large $\mu=6$}
\psfrag{r=1}{\large $\rho=1$}
\psfrag{r=2}{\large $\rho=2$}
\psfrag{Sm=6}{\large $\sum \mu = 6$}
\psfrag{X1}{$\X_1$}
\psfrag{X2}{$\X_2$}
\psfrag{X3}{$\X_3$}
\psfrag{X4}{$\X_4$}
\psfrag{X5}{$\X_5$}
\psfrag{0}{$0$}
\psfrag{1}{$1$}
\psfrag{2}{$2$}
\psfrag{3}{$3$}
\psfrag{4}{$4$}
\psfrag{8}{$8$}
\psfrag{A}{$A$}
\psfrag{B}{$B$}
\psfrag{C}{$C$}
\psfrag{D}{$D$}
\psfrag{N}{$N$}
\psfrag{ld}{$\ldots$}
\psfrag{li}{$l_i$}
\psfrag{ci}{$c_i$}

\def\mid{{\mathop{\rm mid}}}
\def\midt{{\mathop{\rm mid}}}
\def\hi{{\mathop{\rm hi}}}
\def\hit{{\mathop{\rm hi}}}
\def\lo{{\mathop{\rm lo}}}

\def\os{{\mathop{\rm CC}}}
\def\mod{\mathop{\rm mod}}
\def\argmin{\mathop{\arg\,\min}}
\def\power{{\mathop{\rm pow}}}

\def\s{\mbox{'s}}
\def\for{\mbox{ for }}

\def\nodeset{\eta}
\def\Cost{\tilde{L}}
\def\CampCost{L}
\def\CostConst{L_0}
\def\definedas{{}\stackrel {{\mbox{\tiny $\Delta$}}}{{\mbox{\tiny $=$}}}{}}
\def\boldl{{\mbox{\boldmath $l$}}}
\def\smalll{{\mbox{\scriptsize \boldmath $l$}}}

\def\letterl{l}
\def\order{O}

\def\definedas{\triangleq}
\def\das{\triangleq}

\def\posinf{\infty}
\def\lg{{\log_2}}
\def\boldp{{\mbox{\boldmath $p$}}}
\def\p{{\mbox{\boldmath $p$}}}

\def\A{{\mathcal A}}   
\def\I{{\mathcal I}}
\def\letterk{{\kappa}}
\def\N{{\mathbb Z}_+}

\def\R{{\mathbb R}}
\def\Rp{{\mathbb R}_+}

\def\X{{\mathcal X}}
\def\Z{{\mathbb Z}}
\def\gma{{\frac{\alpha}{2\beta}}}
\def\gmat{{\frac{\alpha}{\beta}}}
\def\gmap{{\left({\frac{\alpha}{2\beta}}\right)}}

\def\boldlf{{\boldl^{(f)}}}
\def\boldlg{{\boldl^{(g)}}}
\def\boldlupN{{\boldl^{N}}}
\def\boldlnaa{{\boldl^{(n_1^1)}}}
\def\boldlnbb{{\boldl^{(n_2^2)}}}
\def\boldlncc{{\boldl^{(n_3^3)}}}
\def\boldlna{{\boldl^{(n_1)}}}
\def\boldlnb{{\boldl^{(n_2)}}}
\def\boldlnc{{\boldl^{(n_3)}}}
\def\boldln{{\boldl^{(n)}}}
\def\boldla{{\boldl^{(1)}}}
\def\boldlb{{\boldl^{(2)}}}
\def\boldlc{{\boldl^{(3)}}}
\def\ptn{{\boldp^{(n)}}}

\newcommand{\defn}[0]{\it}

\begin{document}
\title{Source Coding for Quasiarithmetic Penalties}
\author{Michael~B.~Baer,~\IEEEmembership{Member,~IEEE}%
\thanks{This work was supported in part by the National Science Foundation (NSF) under Grant CCR-9973134 and the Multidisciplinary University Research Initiative (MURI) under Grant DAAD-19-99-1-0215.  Material in this paper was presented at the 2003 International Symposium on Information Theory, Yokohama, Japan.}%
\thanks{M. Baer was with the Department of Electrical Engineering,
  Stanford University, Stanford, CA  94305-9505 USA.  He is now with
  Electronics for Imaging, 303 Velocity Way, Foster City, CA  94404
  USA (e-mail: Michael.Baer@efi.com).}
\thanks{This work has been submitted to the IEEE for possible publication. Copyright may be transferred without notice, after which this version may no longer be accessible.}}
\markboth{IEEE Transactions on Information Theory}{Source Coding for Quasiarithmetic Penalties}
\pubid{0000--0000/00\$00.00~\copyright~2006 IEEE}
\maketitle

\begin{abstract}
Huffman coding finds a prefix code that minimizes mean codeword length
for a given probability distribution over a finite number of items.
Campbell generalized the Huffman problem to a family of problems in
which the goal is to minimize not mean codeword length $\sum_i p_i
l_i$ but rather a generalized mean of the form $\varphi^{-1}(\sum_i
p_i \varphi(l_i))$, where $l_i$ denotes the length of the $i$th
codeword, $p_i$ denotes the corresponding probability, and $\varphi$
is a monotonically increasing cost function. Such generalized means
--- also known as quasiarithmetic or quasilinear means --- have a
number of diverse applications, including applications in
queueing. Several quasiarithmetic-mean problems have novel simple
redundancy bounds in terms of a generalized entropy. A related
property involves the existence of optimal codes: For ``well-behaved''
cost functions, optimal codes always exist for (possibly
infinite-alphabet) sources having finite generalized entropy.  Solving
finite instances of such problems is done by generalizing an algorithm
for finding length-limited binary codes to a new algorithm for finding
optimal binary codes for any quasiarithmetic mean with a convex cost
function. This algorithm can be performed using quadratic time and
linear space, and can be extended to other penalty functions, some of
which are solvable with similar space and time complexity, and others
of which are solvable with slightly greater complexity. This reduces
the computational complexity of a problem involving minimum delay in a
queue, allows combinations of previously considered problems to be
optimized, and greatly expands the space of problems solvable in
quadratic time and linear space. The algorithm can be extended for
purposes such as breaking ties among possibly different optimal codes,
as with bottom-merge Huffman coding.
\end{abstract}


\IEEEpeerreviewmaketitle

\section{Introduction} 
\label{intro} 

\PARstart{I}{t} is well known that Huffman coding~\cite{Huff} yields a
prefix code minimizing expected length for a known finite probability
mass function.  Less well known are the many variants of this
algorithm that have been proposed for related problems\cite{Abr01}.
For example, in his doctoral dissertation, Humblet discussed two
problems in queueing that have nonlinear terms to
minimize\cite{Humb0}.  These problems, and many others, can be reduced
to a certain family of generalizations of the Huffman problem
introduced by Campbell in \cite{Camp}.

In all such source coding problems, a source emits symbols drawn from
the alphabet $\X = \{ 1, 2, \ldots, n \}$, where $n$ is an integer (or
possibly infinity).  Symbol $i$ has probability $p_i$, thus defining
probability mass function~$\p$.  We assume without loss of generality
that $p_i > 0$ for every $i \in \X$, and that $p_i \leq p_j$ for every
$i>j$ ($i,j \in \X$).  The source symbols are coded into codewords
composed of symbols of the $\infd$-ary alphabet $\{0, 1, \ldots,
\infd-1\}$, most often the binary alphabet, $\{0,1\}$.  The codeword
$c_i$ corresponding to symbol $i$ has length $\letterl_i$, thus
defining length distribution $\boldl$.  Finding values for $\boldl$ is
sufficient to find a corresponding code.  

Huffman coding minimizes $\sum_{i \in \X} p_i l_i$.  Campbell's
formulation adds a continuous (strictly) monotonic increasing {\defn
cost function} $\varphi(l):\Rp \rightarrow \Rp$.  The value to
minimize is then
\begin{equation}
\CampCost(\boldp,\boldl,\varphi) \definedas \varphi^{-1}\left(\sum_{i \in \X}
p_i \varphi(l_i)\right). \label{CampCost}
\end{equation}
Campbell called (\ref{CampCost}) the ``mean length for the cost
function $\varphi$''; for brevity, we refer to it, or any value to
minimize, as the {\defn penalty}.  Penalties of the form
(\ref{CampCost}) are called {\defn quasiarithmetic} or {\defn
quasilinear}; we use the former term in order to avoid confusion with
the more common use of the latter term in convex optimization theory.

Note that such problems can be mathematically described if we make the natural
coding constraints explicit: the integer constraint, $\letterl_i \in
\Z_+$, and the Kraft (McMillan) inequality~\cite{McMi},
$$ \letterk(\boldl) \definedas \sum_{i \in \X} \infd^{-\letterl_i} \leq 1 . $$

Given these constraints, examples of $\varphi$ in
(\ref{CampCost}) include a quadratic cost function useful in minimizing delay due
to queueing and transmission, 
\begin{equation}
\varphi(x) = \alpha x + \beta x^2 \label{quadratic}
\end{equation}
for nonnegative $\alpha$ and $\beta$\cite{Larm}, and an exponential
cost function useful in minimizing probability of buffer overflow,
$\varphi(x) = \infd^{tx}$ for positive $t$\cite{Humb0,Humb2}.  These
and other examples are reviewed in the next section.

Campbell noted certain properties for convex $\varphi$, such as those
examples above, and others for concave $\varphi$.  Strictly concave
$\varphi$ penalize shorter codewords more harshly than the linear
function and penalize longer codewords less harshly.  Conversely,
strictly convex $\varphi$ penalize longer codewords more harshly than
the linear function and penalize shorter codewords less harshly.
Convex $\varphi$ need not yield convex $\CampCost$, although
$\varphi(\CampCost)$ is clearly convex if and only if $\varphi$ is.
Note that one can map decreasing $\varphi$ to a corresponding
increasing function $\tilde{\varphi}(l) = \varphi_{\max} - \varphi(l)$
without changing the value of $\CampCost$ (e.g., for $\varphi_{\max} = \varphi(0)$).
Thus the restriction to increasing $\varphi$ can be trivially relaxed.


We can generalize $\CampCost$ by using a two-argument cost function
$f(l,p)$ instead of $\varphi(l)$, as in (\ref{Cost}), and
adding $\{\posinf\}$ to its range.  We usually choose functions with
the following property:

\begin{definition}
A cost function $f(l,p)$ and its associated penalty $\Cost$ are 
{\defn differentially monotonic} if, for every $l>1$, whenever
$f(l-1,p_i)$ is finite and $p_i > p_j$,
$f(l,p_i)-f(l-1,p_i) > f(l,p_j)-f(l-1,p_j)$.
\label{difmon}
\end{definition}

\noindent This property means that the contribution to the penalty of
an $l$th bit in a codeword will be greater if the corresponding event
is more likely.  Clearly any $f(l,p)=p\varphi(l)$ will be
differentially monotonic.  This restriction on the generalization will aid in finding
algorithms for coding such cost functions, which we denote as {\defn
generalized quasiarithmetic} penalties:

\begin{definition}
Let $f(l,p):\Rp \times [0,1] \rightarrow \Rp
\cup \{\infty\}$ be a function nondecreasing in $l$.  Then 
\begin{equation}
\Cost(\boldp,\boldl,f) \definedas \sum_{i \in \X} f(l_i, p_i) . \label{Cost}
\end{equation}
is called a {\defn generalized quasiarithmetic} penalty.
Further, if $f$ is convex in $l$, it is called a {\defn
generalized quasiarithmetic convex} penalty.
\end{definition}
As indicated, quasiarithmetic penalties --- mapped with $\varphi$ using $f(l_i,p_i)=p_i
\varphi(l_i)$ to $\Cost(\boldp,\boldl,f) =
\varphi(\CampCost(\boldp,\boldl,\varphi))$ --- are differentially
monotonic, and thus can be considered a special case of differentially
monotonic generalized quasiarithmetic penalties.

In this paper, we seek properties of and algorithms for solving
problems of this form, occasionally with some restrictions (e.g., to
convexity of $\varphi$).  In the next section, we provide examples of
the problem in question.  In Section~\ref{props}, we investigate
Campbell's quasiarithmetic penalties, expanding beyond Campbell's
properties for a certain class of $\varphi$ that we call {\defn
subtranslatory}.  This will extend properties --- entropy bounds,
existence of optimal codes --- previously known only for linear
$\varphi$ and, in the case of entropy bounds, for $\varphi$ of the
exponential form $\varphi(x) = \infd^{tx}$.  These properties pertain
both to finite and infinite input alphabets, and some are applicable
beyond subtranslatory penalties.  We then turn to algorithms for
finding an optimal code for finite alphabets in Section~\ref{alg}; we
start by presenting and extending an alternative to code tree
notation, nodeset notation, originally introduced in \cite{Larm}.
Along with the Coin Collector's problem, this notation can aid in
solving coding problems with generalized quasiarithmetic convex
penalties.  We explain, prove, and refine the resulting algorithm,
which is $\order(n^2)$ time and $\order(n)$ space when minimizing for
a differentially monotonic generalized quasiarithmetic penalty; the
algorithm can be extended to other penalties with a like or slightly
greater complexity.  This is an improvement, for example, on a result
of Larmore, who in \cite{Larm} presented an $\order(n^3)$-time
$\order(n^3)$-space algorithm for cost function (\ref{quadratic}) in
order to optimize a more complicated penalty related to communications
delay.  Our result thus improves overall performance for the quadratic
problem and offers an efficient solution for the more general convex
quasiarithmetic problem.  Conclusions are presented in
Section~\ref{conclusion}.

\section{Examples} 
\label{MaE}

The additive convex coding problem considered here is quite
broad.  Examples include
$$ f(l_i, p_i) = p_i l_i^a \qquad \left(\varphi(x) = x^a\right) $$
for $a \geq 1$, the moment penalty; see, e.g., \cite[pp.~121--122]{Kapu}.
Although efficient solutions have been given for $a=1$ (the
Huffman case) and $a=2$ (the quadratic moment), no polynomial-time
algorithms have been proposed for the general case.

The quadratic moment was considered by Larmore in \cite{Larm} as a
special case of the quadratic problem (\ref{quadratic}), which is
perhaps the case of greatest relevance.  Restating this
problem in terms of $f$,
$$f(l_i, p_i) = p_i (\alpha x + \beta x^2) \qquad \left(\varphi(x) =
\alpha x + \beta x^2\right).$$ This was solved with cubic space and
time complexity as a step in solving a problem related to message
delay.  This larger problem, treated first by Humblet\cite{Humb0}
then Flores\cite{Flor}, was solved with an $\order(n^5)$-time
$\order(n^3)$-space algorithm that can be altered to become an
$\order(n^4)$-time $\order(n^2)$-space algorithm using methods
in this paper.

Another quasiarithmetic penalty is the exponential penalty, that
brought about by the cost function
\begin{equation}
f(l_i, p_i) = p_i \infd^{tl_i} \qquad \left(\varphi(x) = \infd^{tx}\right) \label{expon}
\end{equation}
for $t>0$, $\infd$ being the size of the output alphabet.  This was
previously proposed by Campbell\cite{Camp} and algorithmically solved
as an extension of Huffman's algorithm (and thus with linear time and
space for sorted probability inputs) in \cite{Humb0,HKT,Humb2,Park}.
As previously indicated, in \cite{Humb0,Humb2} this is a step in
minimizing the probability of buffer overflow in a queueing system.
Thus the quasiarithmetic framework includes the two queueing-related
source coding problems discussed in \cite{Humb0}.

A related problem is that with the concave cost function
$$f(l_i, p_i) = p_i (1-\infd^{tl_i}) \qquad \left(\varphi(x) = 1-\infd^{tx}\right)$$
for $t<0$, which has a similar solution\cite{Humb2}.  This problem
relates to a problem in \cite{BaerI06} which is based on a
scenario presented by R\'{e}nyi in \cite{Reny}.

Whereas all of the above, being continuous in $l_i$ and linear in~$p_i$, are
within the class of cases considered by Campbell, the following convex
problem is not, in that its range includes infinity.  Suppose we want
the best code possible with the constraint that all codes must fit
into a structure with $l_{\max}$ symbols.  If our measure of the
``best code'' is linear, then the appropriate penalty is
\begin{equation}
f(l_i, p_i) = \left\{\begin{array}{ll}
p_i l_i,&l_i \leq l_{\max}\\
\posinf,&l_i > l_{\max} \\
\end{array}
\right. \label{llhuff}
\end{equation}
for some fixed $l_{\max} \geq \lceil \log_\infd n \rceil$.  This
describes the length-limited linear penalty, algorithmically solved
efficiently using the Package-Merge algorithm in \cite{LaHi} (with the
assumption that $\infd = 2$).  This approach will be a special
case of our coding algorithm.

Note that if the measure of a ``best code'' is nonlinear, a
combination of penalties should be used where length is limited.  For
example, if we wish to minimize the probability of buffer overflow in
a queueing system with a limited length constraint, we should
combine~(\ref{expon}) and~(\ref{llhuff}):
\begin{equation}
f(l_i, p_i) = \left\{\begin{array}{ll}
p_i \infd^{tl_i},&l_i \leq l_{\max}\\
\posinf,&l_i > l_{\max} \\
\end{array}
\right. 
\label{expll}
\end{equation}
This problem can be solved via dynamic programming in a manner similar
to \cite{Gare}, but this approach takes $\Omega(n^2 l_{\max})$ time
and $\Omega(n^2)$ space for $\infd=\algd$ and greater complexity for
$\infd>\algd$ \cite{Itai}.  Our approach improves on this considerably.

In addition to the above problems with previously known applications
--- and penalties which result from combining these problems --- one
might want to solve for a different utility function in order to find
a compromise among these applications or another trade-off of codeword
lengths.  These functions need not be like Campbell's in that they
need not be linear in $\boldp$; for example, consider $$f(l_i, p_i) =
(1-p_i)^{-l_i}.$$ Although the author knows of no use for this
particular cost function, it is notable as corresponding to one of the
simplest convex-cost penalties of the form (\ref{Cost}).
				   
\section{Properties}
\label{props}

\subsection{Bounds and the Subtranslatory Property} 

Campbell's quasiarithmetic penalty formulation can be restated as follows:
\begin{equation}
\begin{array}{ll}
\mbox{Given } & \p = (p_1, \ldots, p_n),~p_i > 0, \\
& \sum_i p_i = 1 ;\\
& \mbox{convex, monotonically increasing } \\
& \varphi: \Rp \rightarrow \Rp \\
\mbox{Minimize} \,_{\{\smalll\}} &
\CampCost(\boldp,\boldl,\varphi) =
\sum_i p_i \varphi(l_i) \\
\mbox{subject to } & \sum_i \algd^{-l_i} \leq 1; \\
& l_i \in \Z_+
\end{array} 
\label{arith}
\end{equation}
In the case of linear $\varphi$, the integer constraint is often
removed to obtain bounds related to entropy, as we do in the nonlinear
case:
\begin{equation}
\begin{array}{ll}
\mbox{Given } & \p = (p_1, \ldots, p_n),~p_i > 0, \\
& \sum_i p_i = 1 ;\\
& \mbox{convex, monotonically increasing } \\
& \varphi: \Rp \rightarrow \Rp \\
\mbox{Minimize} \,_{\{\smalll\}} &
\CampCost(\boldp,\boldl,\varphi) =
\sum_i p_i \varphi(l_i) \\
\mbox{subject to } & \sum_i \algd^{-l_i} \leq 1 \\
\end{array} \label{ideal}
\end{equation}
Note that, given $\p$ and $\varphi$, $\CampCost^\dagger$, the minimum
for the relaxed (real-valued) problem (\ref{ideal}), will necessarily
be less than or equal to $\CampCost^*$, the minimum for the original
(integer-constrained) problem (\ref{arith}).  Let $\boldl^\dagger$ and
$\boldl^*$ be corresponding minimizing values for the relaxed and
constrained problems, respectively.  Restating, and adding a fifth
definition:
\begin{eqnarray*}
\CampCost^* &\definedas&
\min_
{\stackrel
{ \sum_i \infd^{-l_i} \leq 1,}
{l_i \in \Z_+}
} \CampCost(\boldp,\boldl,\varphi) \\
\boldl^* &\definedas&
\argmin_
{\stackrel
{ \sum_i \infd^{-l_i} \leq 1,}
{l_i \in \Z_+}
} \CampCost(\boldp,\boldl,\varphi) \\
\CampCost^\dagger &\definedas&
\min_
{\stackrel
{ \sum_i \infd^{-l_i} \leq 1,}
{l_i \in \R}
} \CampCost(\boldp,\boldl,\varphi) \\
\boldl^\dagger &\definedas&
\argmin_
{\stackrel
{ \sum_i \infd^{-l_i} \leq 1,}
{l_i \in \R}
} \CampCost(\boldp,\boldl,\varphi) \\
\boldl^\ddagger &\definedas&
(\lceil l_1^\dagger \rceil, \lceil l_2^\dagger \rceil, \ldots, \lceil
l_n^\dagger \rceil)
\end{eqnarray*}
This is a slight abuse of $\argmin$ notation since $\CampCost^*$ could
have multiple corresponding optimal length distributions ($\boldl^*$).
However, this is not a problem, as any such value will suffice.  Note
too that $\boldl^\ddagger$ satisfies the Kraft inequality and the integer
constraint, and thus $\CampCost(\boldp,\boldl^\ddagger,\varphi) \geq
\CampCost^*$.

We obtain bounds for the optimal solution by noting that,
since $\varphi$ is monotonically increasing,
\begin{equation}
\begin{array}{rcl}
\varphi^{-1}\left(\sum_i p_i \varphi(l_i^\dagger)\right) & \leq &
\varphi^{-1}\left(\sum_i p_i \varphi(l_i^*)\right) \\
& \leq & \varphi^{-1}\left(\sum_i p_i \varphi(l_i^\ddagger)\right) \\
& < & \varphi^{-1}\left(\sum_i p_i \varphi(l_i^\dagger + 1)\right).
\end{array} \label{bounds}
\end{equation}

These bounds are similar to Shannon redundancy bounds for Huffman
coding.  In the linear/Shannon case, $l_i^\dagger = - \lg p_i$, so the
last expression is $\sum_i p_i (l_i^\dagger + 1) = 1+\sum_i p_i
l_i^\dagger = 1 + H(\p)$, where $H(\p)$ is the Shannon entropy, so
$H(\p) \leq \sum_i p_il_i^* < 1+H(\p)$.  These Shannon bounds can be
extended to quasiarithmetic problems by first defining
$\varphi$-entropy as follows:

\begin{definition}
{\defn Generalized entropy} or {\defn $\varphi$-entropy} is
\begin{equation}
H(\p, \varphi) \definedas 
\inf_
{\stackrel
{ \sum_i \infd^{-l_i} \leq 1,}
{l_i \in \R}
}
\CampCost(\boldp,\boldl,\varphi) 
\label{entropy}
\end{equation}
where here infimum is used because this definition applies to codes with 
infinite, as well as finite, input alphabets\cite{Camp}.
\end{definition}

Campbell defined this as a generalized entropy \cite{Camp}; we go further, by
asking which cost functions, $\varphi$, have the following property:
\begin{equation}
H(\p,\varphi) \leq \CampCost(\boldp, \boldl^*,\varphi) < 1+H(\p, \varphi) \label{Fbound}
\end{equation}

These bounds exist for the exponential case (\ref{expon}) with
$H(\p,\varphi) = H_\alpha(\p)$, where $\alpha \das (1+t)^{-1}$,
and $H_\alpha(\p)$ denotes R\'{e}nyi $\alpha$-entropy\cite{Ren1}.  The
bounds extend to exponential costs because they share with the linear
costs (and only those costs) a property known as the {\defn
translatory} property, described by Acz\'{e}l\cite{Acz3}, among
others:

\begin{definition} 
A cost function $\varphi$ (and its associated penalty) is {\defn
translatory} if, for any $\boldl \in \Rp^n$, probability mass function~$\p$,
and $c \in \Rp$, 
$$\CampCost(\boldp,\boldl+c,\varphi) = \CampCost(\boldp,\boldl,\varphi)+c$$
where $\boldl+c$ denotes adding
$c$ to each $l_i$ in $\boldl$ \cite{Acz3}.
\end{definition}

We broaden the collection of penalty functions satisfying such bounds by
replacing the translatory equality with an
inequality, introducing the concept of a {\defn subtranslatory}
penalty:

\begin{definition} 
A cost function $\varphi$ (and its associated penalty) is {\defn
  subtranslatory} if, for any $\boldl \in \Rp^n$, probability mass function
  $\p$, and $c \in \Rp$,
$$\CampCost(\boldp,\boldl+c,\varphi) \leq \CampCost(\boldp,\boldl,\varphi)+c.$$
\end{definition}
For such a penalty, (\ref{Fbound}) still holds.

If $\varphi$ obeys certain regularity requirements, then we can introduce
a necessary and sufficient condition for it to be subtranslatory.
Suppose that the invertible function $\varphi: \Rp \rightarrow \Rp$
is real analytic over a relevant compact interval.  We might
choose this interval to be, for example, $\A = [\delta, 1/\delta]$ for
some $\delta \in (0,1)$.  (Let $\delta \rightarrow 0$ to show the
following argument is valid over all $\Rp$.)  We assume $\varphi^{-1}$ is
also real analytic (with respect to interval $\varphi(\A)$).  Thus all
derivatives of the function and its inverse are bounded.  

\begin{theorem}
Given real analytic cost function $\varphi$ and its real analytic
inverse $\varphi^{-1}$, $\varphi$ is subtranslatory if and only if,
for all positive $\boldl$ and all positive $\p$ summing to~$1$, 
\begin{equation}
\sum_i p_i \varphi'(l_i) \leq \varphi'\left(\varphi^{-1}\left(\sum_i p_i
  \varphi(l_i)\right)\right) \label{transsuff}
\end{equation}
where $\varphi'$ is the derivative of $\varphi$.
\end{theorem}

\begin{proof}
First note that, since all values are positive, inequality
(\ref{transsuff}) is equivalent to
\begin{equation}
\left(\sum_i p_i \varphi'(l_i)\right) \cdot
(\varphi^{-1})'\left(\sum_i p_i \varphi(l_i)\right) \leq 1 .
\label{theq}
\end{equation}

We show that, when (\ref{theq}) is true everywhere, $\varphi$ is
subtranslatory, and then we show the converse.  Let $\epsilon > 0$.
Using power expansions of the form $$g(x) + \epsilon g'(x) =
g(x+\epsilon) \pm \order(\epsilon^2)$$ on $\varphi$ and $\varphi^{-1}$,
\begin{equation}
\begin{array}{l}
\displaystyle
\varphi^{-1}\left(\sum_i p_i \varphi(l_i)\right) + \epsilon \\
\displaystyle \quad
\stackrel{{\mbox{\tiny (a)}}}{{\mbox{\tiny $\geq$}}}{}
\varphi^{-1}\left(\sum_i p_i \varphi(l_i)\right) \\
\displaystyle \qquad
+~\epsilon \cdot \left(\sum_i p_i \varphi'(l_i)\right) \cdot
(\varphi^{-1})'\left(\sum_i p_i \varphi(l_i)\right) \\
\displaystyle \quad 
\stackrel{{\mbox{\tiny (b)}}}{{\mbox{\tiny $=$}}}{}
 \varphi^{-1}\left(\sum_i p_i \varphi(l_i) + \epsilon \cdot \sum_i p_i
\varphi'(l_i)\right) \pm \order(\epsilon^2) \\
\displaystyle \quad 
\stackrel{{\mbox{\tiny (c)}}}{{\mbox{\tiny $=$}}}{}
\varphi^{-1}\left(\sum_i p_i \varphi(l_i+\epsilon) \pm \order(\epsilon^2)\right)
\pm \order(\epsilon^2) \\
\displaystyle \quad 
\stackrel{{\mbox{\tiny (d)}}}{{\mbox{\tiny $=$}}}{}
 \varphi^{-1}\left(\sum_i p_i \varphi(l_i+\epsilon)\right) \pm \order(\epsilon^2). 
\end{array}
\label{phiphi}
\end{equation}
Step (a) is due to (\ref{theq}), step (b) due to the power
expansion on $\varphi^{-1}$, step (c) due to the power expansion on
$\varphi$, and step (d) due to the power expansion on $\varphi^{-1}$ (where
the bounded derivative of $\varphi^{-1}$ allows for the asymptotic
term to be brought outside the function).

Next, evoke the above inequality $c/\epsilon$ times:
\begin{equation}
\begin{array}{l}
\displaystyle
\varphi^{-1}\left(\sum_i p_i \varphi(l_i+c)\right) \\ 
\displaystyle
\quad \leq \epsilon + \varphi^{-1}\left(\sum_i p_i \varphi(l_i + c - \epsilon)\right)
\pm \order(\epsilon^2) \\
\displaystyle
\quad \leq \cdots \\
\displaystyle
\quad \leq \epsilon \left\lfloor \frac{c}{\epsilon} \right\rfloor + \varphi^{-1}\left(\sum_i p_i
  \varphi(l_i) + c - \epsilon \left\lfloor \frac{c}{\epsilon} \right\rfloor\right) \pm \order(\epsilon) \\
\displaystyle
\quad \leq c + \varphi^{-1}\left(\sum_i p_i \varphi(l_i)\right) \pm \order(\epsilon)
\end{array}
\label{epsineq}
\end{equation}
Taking $\epsilon \rightarrow 0$, 
$$\varphi^{-1}\left(\sum_i p_i \varphi(l_i+c)\right) \leq c+\varphi^{-1}\left(\sum_i
p_i \varphi(l_i)\right).$$ Thus, the fact of (\ref{transsuff}) is
sufficient to know that the penalty is subtranslatory.

To prove the converse, suppose $\sum_i p_i \varphi'(l_i) >
\varphi'\left(\varphi^{-1}\left(\sum_i p_i \varphi(l_i)\right)\right)$
for some valid $\boldl$ and $\boldp$.  Because $\varphi$ is analytic,
continuity implies that there exist $\delta_0 > 0$ and $\epsilon_0 >
0$ such that 
$$ \sum_i p_i \varphi'(l'_i) \geq (1+\delta_0) \cdot
\varphi'\left(\varphi^{-1}\left(\sum_i p_i
\varphi(l'_i)\right)\right)$$ for all $\boldl' \in
[\boldl,\boldl+\epsilon_0)$.  The chain of inequalities above reverse
in this range with the additional multiplicative constant.  Thus
(\ref{phiphi}) becomes
$$
\begin{array}{l}
{\displaystyle
\varphi^{-1}\left(\sum_i p_i \varphi(l'_i)\right) + (1+\delta_0)\epsilon} \\
{\displaystyle
\quad \leq
\varphi^{-1}\left(\sum_i p_i \varphi(l'_i+\epsilon)\right) 
\pm \order(\epsilon^2)}
\end{array}
$$ 
for $\boldl' \in [\boldl,\boldl+\epsilon_0)$, and (\ref{epsineq}) becomes, for any $c \in
(0,\epsilon_0)$,
$$
\begin{array}{l}
{\displaystyle
\varphi^{-1}\left(\sum_i p_i \varphi(l_i+c)\right) } \\ 
{\displaystyle
\quad \geq (1+\delta_0)c + \varphi^{-1}\left(\sum_i p_i \varphi(l_i)\right) \pm \order(\epsilon)}
\end{array}
$$
which, taking $\epsilon \rightarrow 0$, similarly leads to
$$
\begin{array}{l}
{\displaystyle
\varphi^{-1}\left(\sum_i p_i \varphi(l_i+c)\right)} \\
{\displaystyle
\quad \geq (1+\delta_0)c + \varphi^{-1}\left(\sum_i p_i
\varphi(l_i)\right)} \\
{\displaystyle
\quad > c + \varphi^{-1}\left(\sum_i p_i \varphi(l_i)\right)}
\end{array}
$$
and thus the subtranslatory property fails and the
converse is proved.
\end{proof} 

Therefore, for $\varphi$ satisfying (\ref{transsuff}), we have the
bounds of (\ref{Fbound}) for the optimum solution.  Note that the
right-hand side of (\ref{transsuff}) may also be written
$\varphi'\left(\CampCost(\p,\boldl,\varphi)\right)$; thus
(\ref{transsuff}) indicates that the average derivative of $\varphi$
at the codeword length values is at most the derivative of $\varphi$
at the value of the penalty for those length values.

The linear and exponential penalties satisfy these equivalent
inequalities with equality.  Another family of cost functions that
satisfies the subtranslatory property is $\varphi(l_i)=l_i^a$
for fixed $a \geq 1$, which corresponds to
$$\CampCost(\p,\boldl,\varphi)=\left(\sum_i p_i l_i^a\right)^{1/a}.$$
Proving this involves noting that Lyapunov's inequality for moments of
a random variable yields
$$\left(\sum_i p_i l_i^{a-1} \right)^\frac{1}{a-1}
\leq \left(\sum_i p_i l_i^{a} \right)^\frac{1}{a}$$
which leads to 
$$a \cdot \left(\sum_i p_i l_i^{a-1} \right) \leq a \cdot \left(\sum_i
p_i l_i^{a-1} \right)^\frac{a-1}{a}$$ which, because
$\varphi'(x)=ax^{a-1}$, is
$$\sum_i p_i \varphi'(l_i) \leq \varphi'\left(\varphi^{-1}\left(\sum_i p_i \varphi(l_i)\right)\right)$$ 
the inequality we desire.

Another subtranslatory penalty is the quadratic quasiarithmetic penalty of
(\ref{quadratic}), in which $$\varphi(x) = \alpha x + \beta x^2$$ for
$\alpha, \beta \geq 0$.  This has already been shown for $\beta = 0$;
when $\beta > 0$, 
\begin{eqnarray*}
\varphi'(x) &=& \alpha + 2 \beta x \\
\varphi^{-1}(x) &=& \sqrt{\gmap ^2 - \frac{x}{\beta}} - \gma \\
\CampCost(\p,\boldl,\varphi) &=& \sqrt{\gmap ^2+\sum_i p_i \left(\gmat l_i + l_i^2\right)} -
\gma.
\end{eqnarray*}
We achieve the desired inequality through algebra:
\begin{eqnarray*}
\sum_i p_i l_i^2 
& \geq & 
\left(\sum_i p_i l_i\right)^2 \\
\alpha^2 + 4 \beta \sum_i p_i (\alpha l_i + \beta
  l_i^2) & \geq &
\left(\sum_i p_i (\alpha + 2 \beta l_i)\right)^2 \\
\sqrt{\alpha^2 + 4 \beta \sum_i p_i (\alpha l_i + \beta
  l_i^2)} & \geq &
\sum_i p_i (\alpha + 2 \beta l_i) \\
\varphi'(\CampCost(\p,\boldl,\varphi)) & \geq & \sum_i p_i \varphi'(l_i)
\end{eqnarray*}
We thus have an important property that holds for several cases of interest.

One might be tempted to conclude that every $\varphi$ --- or every
convex and/or concave $\varphi$ --- is subtranslatory.  However, this is easily
disproved.  Consider convex $\varphi(x) = x^3 + 11x$.  Using Cardano's
formula, it is easily seen that (\ref{transsuff}) does not
hold for $\p=(\frac{1}{3},\frac{2}{3})$ and $\boldl = (\frac{1}{2},
1)$.  The subtranslatory test also fails for $\varphi(x) = \sqrt{x}$.
Thus we must test any given penalty for the subtranslatory property in
order to use the redundancy bounds.

\subsection{Existence of an Optimal Code}

Because all costs are positive, the redundancy bounds that are a
result of a subtranslatory penalty extend to infinite alphabet codes
in a straightforward manner.  These bounds thus show that a code with
finite penalty exists if and only if the generalized entropy is
finite, a property we extend to nonsubtranslatory penalties in
the next subsection.  However, one must be careful regarding the meaning of
an ``optimal code'' when there are an infinite number of possible
codes satisfying the Kraft inequality with equality.  Must there exist
an optimal code, or can there be an infinite sequence of codes of
decreasing penalty without a code achieving the limit penalty value?

Fortunately, the answer is the former, as the existence results of
Linder, Tarokh, and Zeger in \cite{LTZ} can be extended to
quasiarithmetic penalties.  Consider continuous strictly monotonic
$\varphi:\Rp \rightarrow \Rp$ (as proposed by Campbell) and $\p = (
p_1, p_2, \ldots )$ such that
\begin{equation}
\CampCost^*(\p,\varphi) \definedas \inf_
{\stackrel
{ \sum_i \infd^{-l_i} \leq 1,}
{l_i \in \Z}
}
\varphi^{-1} \left( \sum_{i=1}^\posinf p_i \varphi(l_i) \right) 
\label{ccstar}
\end{equation}
is finite.  Consider, for an arbitrary $n \in \N$, optimizing
for $\varphi$ with weights 
$$\ptn \definedas (p_1, p_2, \ldots, p_n, 0, 0, \ldots).$$
(We call the entries to this distribution ``weights'' because they
do not necessarily add up to $1$.)  
Denote the optimal code a {\defn truncated code}, one with codeword lengths
$$\boldln \definedas \{l_1^{(n)}, l_2^{(n)}, \ldots, l_n^{(n)}, \infty, \infty, \ldots\}.
$$
Thus, for convenience, $l_i^{(j)} = \infty$ for $i>j$.  These lengths are
also optimal for $(\sum_{j=1}^n p_j)^{-1} \cdot \ptn$, the
distribution of normalized weights.

Following \cite{LTZ}, we say that a sequence of codeword length
distributions $\boldla, \boldlb, \boldlc, \ldots$ {\defn converges}
to an infinite prefix code with codeword lengths $\boldl = \{l_1,
l_2, \ldots \}$ if, for each $i$, the $i$th length in each
distribution in the sequence is eventually $l_i$ (i.e., if each
sequence converges to $l_i$).

\begin{theorem}
Given quasiarithmetic increasing $\varphi$ and $\p$
such that $\CampCost^*(\p, \varphi)$ is finite, the following hold:

\begin{enumerate}
\item There exists a sequence of truncated codeword lengths that
  converges to optimal codeword lengths for $\p$; thus the infimum is achievable.
\item Any optimal code for $\p$ must satisfy the Kraft inequality
  with equality.
\end{enumerate}
\end{theorem}

\begin{proof}
  Because here we are concerned only with cases in which the first length
  is at least $1$, we may restrict ourselves to the domain $[\varphi^{-1}(p_1
  \varphi(1)),\posinf)$. Recall
$$ \CampCost^*(\p, \varphi) = \inf_{\stackrel
{ \sum_i \infd^{-l_i} \leq 1,}
{l_i \in \Z}
}
\varphi^{-1} \left( \sum_{i=1}^\posinf p_i \varphi(l_i) \right) < \posinf.
$$

Then there exists near-optimal $\boldl' = \{l'_1, l'_2, l'_3, \ldots\} \in \N^\posinf$ such that
$$
\varphi^{-1} \left( \sum_{i=1}^\posinf p_i \varphi(l'_i) \right) < 
\CampCost^*(\p,\varphi)+1 \mbox{ and }
\sum_{i=1}^\posinf \infd^{-l'_i} \leq 1 ,
$$
and thus, for any integer $n$,
$$
\varphi^{-1} \left( \sum_{i=1}^n p_i \varphi(l'_i) \right) < 
\CampCost^*(\p,\varphi)+1 \mbox{ and }
\sum_{i=1}^n \infd^{-l'_i} < 1 . 
$$

So, using this to approximate the behavior of a minimizing
$\boldln$, we have
\begin{eqnarray*}
\varphi^{-1} \left( \sum_{i=1}^n p_i \varphi(l_i^{(n)}) \right) &\leq& 
\varphi^{-1} \left( \sum_{i=1}^n p_i \varphi(l'_i) \right) \\ &<& \CampCost^*(\p,\varphi)+1
\end{eqnarray*}
yielding an upper bound on terms
\begin{eqnarray*}
p_j \varphi(l_j^{(n)}) &\leq& \sum_{i=1}^n p_i \varphi(l_i^{(n)}) \\
&<& \varphi\left(\CampCost^*(\p,\varphi)+1\right)
\end{eqnarray*}
for all $j$.  This implies 
$$ l_j^{(n)} < \varphi^{-1}\left( \frac{\varphi(\CampCost^*(\p,\varphi)+1)}{p_j} \right) . $$

Thus, for any $i \in \N$, the sequence 
$l_i^{(1)},l_i^{(2)}, l_i^{(3)}, \ldots$ is bounded for all $l_i^{(j)} \neq
\infty$, and thus has a finite set of values (including $\infty$).  It
is shown in \cite{LTZ} that this sufficies for the desired convergence, but for completeness a slightly altered proof
follows.  

Because each sequence $l_i^{(1)}, l_i^{(2)}, l_i^{(3)}, \ldots$ has a
finite set of values, every infinite indexed subsequence for a given
$i$ has a convergent subsequence.  An inductive argument implies that,
for any $k$, there exists a subsequence indexed by $n_j^{k}$ such that
$l_i^{(n_1^{k})}, l_i^{(n_2^{k})}, l_i^{(n_3^{k})}, \ldots$ converges
for all $i \leq k$, where $l_i^{(n_1^k)}, l_i^{(n_2^k)},
l_i^{(n_3^k)}, \ldots$ is a subsequence of $l_i^{(n_1^{k'})},
l_i^{(n_2^{k'})}, l_i^{(n_3^{k'})}, \ldots$ for $k' \leq k$.  Codeword
length distributions $\boldlnaa, \boldlnbb, \boldlncc, \ldots$ (which
we call $\boldlna, \boldlnb, \boldlnc, \ldots$) thus converge to the
codeword lengths of an infinite code $\widehat{C}$ with codeword
lengths $\widehat{\boldl} = \{\widehat{l_1}, \widehat{l_2},
\widehat{l_3}, \ldots \}$.  Clearly each codeword length distribution
satisfies the Kraft inequality.  The limit does as well then; were it
exceeded, we could find $i'$ such that
$$\sum_{i=1}^{i'} \infd^{-\widehat{l_i}} > 1$$ and thus $n'$ such that 
$$\sum_{i=1}^{i'} \infd^{-{l_i^{(n')}}} > 1$$ causing a
contradiction.

We now show that $\widehat{C}$ is optimal.  Let $\{\lambda_1, \lambda_2,
\lambda_3 \ldots\}$ be the codeword lengths of an arbitrary prefix
code.  For every $k$, there is a $j \geq k$ such that $\widehat{l_i} =
l_i^{(n_m)}$ for any $i \leq k$ if $m \geq j$.  Due to the optimality of
each $\boldln$, for all $m \geq j$:
\begin{eqnarray*}
\sum_{i=1}^k p_i \varphi(\widehat{l_i}) &=& \sum_{i=1}^k p_i \varphi(l_i^{(n_m)}) \\
&\leq& \sum_{i=1}^{n_m} p_i \varphi(l_i^{(n_m)}) \\
&\leq& \sum_{i=1}^{n_m} p_i
\varphi(\lambda_i) \\
&\leq& \sum_{i=1}^\posinf p_i \varphi(\lambda_i) 
\end{eqnarray*}
and, taking $k \rightarrow \posinf$, $\sum_i p_i
\varphi(\widehat{l_i}) \leq \sum_i p_i \varphi(\lambda_i)$, leading
directly to $\varphi^{-1}\left( \sum_i p_i \varphi(\widehat{l_i})
\right) \leq \varphi^{-1}\left( \sum_i p_i \varphi(\lambda_i) \right)$
and the optimality of $\widehat{C}$.

Suppose the Kraft inequality is
not satisfied with equality for optimal codeword lengths
$\widehat{\boldl} = \{\widehat{l_1}, \widehat{l_2}, \ldots \}$.  We can
then produce a strictly superior code.  There is a $k \in \N$ such
that $\infd^{-l_k+1} + \sum_i \infd^{-l_i} \leq 1$.  Consider code
$\{\widehat{l_1}, \widehat{l_2}, \ldots, \widehat{l_{k-1}},
\widehat{l_k}-1, \widehat{l_{k+1}}, \widehat{l_{k+2}}, \ldots\}$.
This code satisfies the Kraft inequality and has penalty
$\varphi^{-1}\left(\sum_i p_i \varphi(\widehat{l_i}) + p_k
(\varphi(\widehat{l_k}-1) - \varphi(\widehat{l_k}))\right) <
\varphi^{-1}\left(\sum_i p_i \varphi(\widehat{l_i}) \right)$.  Thus
$\widehat{\boldl}$ is not optimal.  Therefore the Kraft inequality must
be satisfied with equality for optimal infinite codes.
\end{proof}

Note that this theorem holds not just for subtranslatory penalties,
but for any quasiarithmetic penalty.

\subsection{Finiteness of Penalty for an Optimal Code}

Recall the definition of (\ref{entropy}),
$$
H(\p,\varphi) = \inf_
{\stackrel
{ \sum_i \infd^{-l_i} \leq 1,}
{l_i \in \R}
}
\varphi^{-1} \left( \sum_i p_i \varphi(l_i) \right) 
$$
for $\varphi: \Rp \rightarrow \Rp$.  

\begin{theorem}
If $H(\p,\varphi)$ is finite and
either $\varphi$ is subtranslatory or $\varphi(x+1) =
\order(\varphi(x))$ (which includes all concave and all polynomial $\varphi$),
then the coding problem of (\ref{ccstar}),
$$
\CampCost^*(\p,\varphi) = 
\inf_
{\stackrel
{ \sum_i \infd^{-l_i} \leq 1,}
{l_i \in \Z}
}
\varphi^{-1} \left( \sum_i p_i \varphi(l_i) \right) 
$$
has a minimizing $\boldl^*$ resulting in a finite value for $\CampCost^*(\p,\varphi)$.
\end{theorem}

\begin{proof}
If $\varphi$ is
subtranslatory, then $\CampCost^*(\p,\varphi) < 1 + H(\p,\varphi) < \posinf$.  If $\varphi(x+1) =
\order(\varphi(x))$, then there are $\alpha, \beta > 0$ such that $\varphi(x+1) <
\alpha+\beta \varphi(x)$ for all $x$.  Then
$$
\begin{array}{l}
\displaystyle
\varphi^{-1} \left( \sum_i p_i \varphi(l_i+1) \right) \\
\displaystyle
\quad < \varphi^{-1} \left( \sum_i p_i (\alpha+\beta \varphi(l_i)) \right) \\ 
\displaystyle
\quad = \varphi^{-1} \left( \alpha + \beta \sum_i p_i \varphi(l_i) \right).
\end{array}
$$
So 
$$
\begin{array}{l}
\displaystyle \CampCost^*(\p,\varphi) \\
\displaystyle \quad < \CampCost(\p,\boldl^\dagger+1, \varphi) \\
\displaystyle \quad < \varphi^{-1} \left( \alpha + \beta \varphi\left(H(\p,\varphi)\right) \right) \\
\quad < \posinf
\end{array}
$$
and the infimum, which we know to also be a minimum, is finite.
\end{proof}

\section{Algorithms}
\label{alg}
\subsection{Nodeset Notation}
\label{nodeset}

We now examine algorithms for finding minimum penalty codes for convex
cases with finite alphabets.  We first present a notation for codes
based on an approach of Larmore \cite{Larm}.  This notation is an
alternative to the well known code tree notation, e.g., \cite{Schw}, and it will be the basis for an algorithm to solve the
generalized quasiarithmetic (and thus Campbell's quasiarithmetic)
convex coding problem.

In the literature nodeset notation is generally used for binary
alphabets, not for general alphabet coding.  Although we briefly
sketch how to adapt this technique to general output alphabet coding
at the end of Subsection~\ref{refine}, an approach fully explained in
\cite{Baer20}, until then we concentrate on the binary case ($\infd =
2$).

{\it The key idea:} Each node $(i,l)$ represents both the share of the
penalty $\Cost(\p, \boldl, f)$ (weight) and the share of the Kraft sum
$\letterk(\boldl)$ (width) assumed for the $l$th bit of the $i$th
codeword.  If we show that total weight is an increasing function of
the penalty and show a one-to-one correspondence between optimal
nodesets and optimal codes, we can reduce the problem to an
efficiently solvable problem, the Coin Collector's problem.

In order to do this, we first assume bounds on the maximum codeword
length of possible solutions, e.g., the maximum unary codeword length
of $n-1$.  Alternatively, bounds might be explicit in the definition
of the problem.  Consider for example the length-limited coding
problems of (\ref{llhuff}) and (\ref{expll}), upper bounded by $l_{\max}$.
A third possibility is that maximum length may be implicit in some
property of the set of optimal solutions\cite{KaNe, Buro, AbMc}; we
explore this in Subsection~\ref{refine}.

\begin{figure*}
\begin{center}
\includegraphics{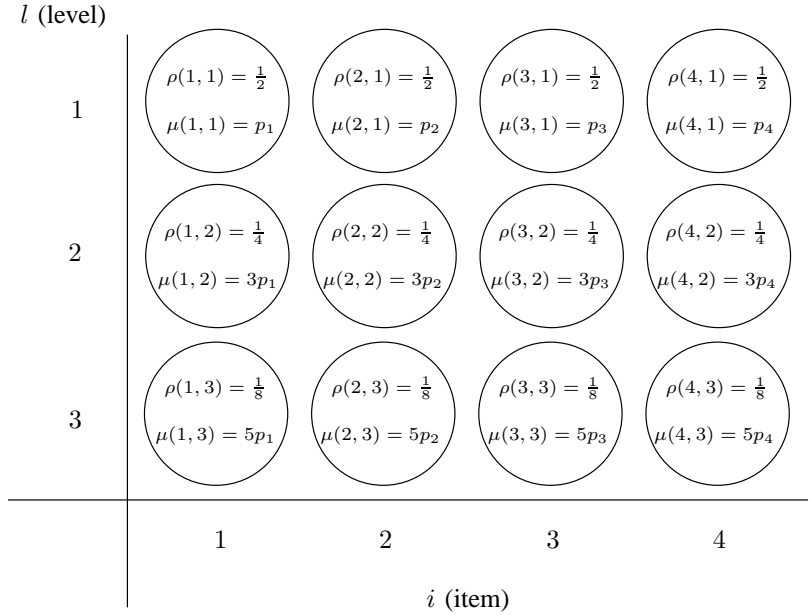}
\caption{The set of nodes $I$ with widths $\{\rho(i,l)\}$ and weights 
$\{\mu(i,l)\}$ for $f(l_i,p_i) = p_il_i^2$, $n=4$, $l_{\max}=3$}
\label{nodesetnum}
\end{center}
\end{figure*}

We therefore restrict ourselves to codes with $n$ codewords, none
of which has greater length than $l_{\max}$, where $l_{\max} \in
[\lceil \log_\algd n \rceil, n-1]$.  With this we now introduce the
{\defn nodeset} notation for binary coding:

\begin{definition} A {\defn node} is an ordered pair of integers $(i, l)$ such
  that $i \in \{1,\ldots,n\}$ and $l \in \{1,\ldots,l_{\max}\}$.
  Call the set of all $nl_{\max}$ possible nodes~$I$.  Usually $I$
  is arranged in a grid; see example in Fig.~\ref{nodesetnum}.  The
  set of nodes, or {\defn nodeset}, corresponding to item $i$
  (assigned codeword~$c_i$ with length~$l_i$) is the set of the
  first~$l_i$ nodes of column~$i$, that is, $\nodeset_\smalll(i)
  \das \{(j,l)~|~j=i,~l \in \{1,\ldots,l_i\}\}$.  The nodeset
  corresponding to length distribution~$\boldl$ is $\nodeset(\boldl)
  \das $ $\bigcup_i \nodeset_\smalll(i)$;
  this corresponds to a set of $n$ codewords, a code.  We say a node
  $(i,l)$ has {\defn width} $\rho(i,l) \das \algd^{-l}$ and {\defn
  weight} $\mu(i,l) \das f(l, p_i) - f(l-1, p_i)$, as in the
  example in Fig.~\ref{nodesetnum}.
\end{definition}

If $I$ has a subset $N$ that is a valid nodeset, then it is
straightforward to find the corresponding length distribution and thus a
code.  We can find an optimal valid nodeset using the Coin Collector's
problem.

\subsection{The Coin Collector's Problem}
\label{cc}
Let $\algd^\Z$ denote the set of all integer powers of two.  The Coin
Collector's problem of size $m$ considers $m$ ``coins'' with width $\rho_i \in
\algd^\Z$; one can think of width as coin face value, e.g., $\rho_i =
\frac{1}{4}$ for a quarter dollar (25 cents).  Each coin also has weight $\mu_i \in \R$.
The final problem parameter is total width, denoted $t$.  The problem is then:
\begin{equation}
\begin{array}{ll}
\mbox{Minimize} \,_{\{B \subseteq \{1,\ldots,m\}\}} & \sum_{i \in B} \mu_i  \\
\mbox{subject to } & \sum_{i \in B} \rho_i = t 
\end{array} \label{knap}
\end{equation}
We thus wish to choose coins with total width $t$ such that their total
weight is as small as possible.  This problem is an input-restricted
variant of the knapsack problem, which, in general, is NP-hard; no
polynomial-time algorithms are known for such NP-hard
problems\cite{GaJo, Manb}.  However, given sorted inputs, a
linear-time solution to (\ref{knap}) was proposed in \cite{LaHi}.  The
algorithm in question is called the {\defn Package-Merge algorithm}.

In the Appendix, we illustrate and prove a slightly simplified version
of the Package-Merge algorithm.  This algorithm allows us to solve the
generalized quasiarithmetic convex coding problem~(\ref{Cost}).  When we
use this algorithm, we let $\I$ represent the $m$ items along with
their weights and widths.  The optimal solution to the problem is a
function of total width $t$ and items $\I$.  We denote this solution
as $\os(\I,t)$ (read, ``the [optimal] coin collection for $\I$ and
$t$'').  Note that, due to ties, this need not be unique, but we
assume that one of the optimal solutions is chosen; at the end of
Subsection~\ref{linear}, we discuss which of the optimal solutions is
best to choose.

\subsection{A General Algorithm}
\label{algorithm}

We now formalize the reduction from the generalized quasiarithmetic convex
coding problem to the Coin Collector's problem.

We assert that any optimal solution $N$ of the Coin
Collector's problem for $t=n-1$ on coins $\I=I$ is a nodeset for an
optimal solution of the coding problem.  This yields a suitable
method for solving generalized quasiarithmetic convex penalties.

To show this reduction, first define $\rho(N)$ for any $N = \nodeset(\boldl)$:
\begin{eqnarray*}
\rho(N) &\definedas & \sum_{(i,l) \in N} \rho(i,l) \\
&=& \sum_{i=1}^n \sum_{l=1}^{l_i} \algd^{-l}\\ 
&=& \sum_{i=1}^n \left(1 -
\algd^{-l_i}\right) \\ 
&=& n - \sum_{i=1}^n \algd^{-l_i} \\
&=& n - \letterk(\boldl)
\end{eqnarray*}
Because the Kraft inequality is $\letterk(\boldl) \leq 1$, $\rho(N)$
must lie in $[n-1,n)$ for prefix codes.  The Kraft inequality is
satisfied with equality at the left end of this interval.  Optimal
binary codes have this equality satisfied, since a strict
inequality implies that the longest codeword length can be shortened
by one, strictly decreasing the penalty without violating the
inequality.  Thus the optimal solution has $\rho(N)=n-1$.

Also define:
\begin{eqnarray*}
M_f(l,p) &\definedas& f(l,p)-f(l-1,p) \\
\CostConst(\p,f) &\definedas& \sum_{i=1}^n f(0,p_i) \\
\mu(N) &\definedas & \sum_{(i,l) \in N} \mu(i,l)
\end{eqnarray*}
Note that
\begin{eqnarray*}
\mu(N) &=& \sum_{(i,l) \in N} \mu(i,l) \\
&=& \sum_{i=1}^n \sum_{l=1}^{l_i} M_f(l, p_i) \\
&=& \sum_{i=1}^n f(l_i, p_i) - \sum_{i=1}^n f(0,p_i)
\\
&=& \Cost(\p,\boldl,f) - \CostConst(\p,f).
\end{eqnarray*}
$\CostConst(\p,f)$ is a constant given fixed penalty and
probability distribution.  Thus, if the optimal nodeset corresponds to
a valid code, solving the Coin Collector's problem solves this coding
problem.  To prove the reduction, we need to prove that the optimal
nodeset indeed corresponds to a valid code.  We begin with the
following lemma:

\begin{lemma}
\label{lllemma}
Suppose that $N$ is a nodeset of width $x\algd^{-k}+r$ where $k$ and $x$ are
integers and $0<r<\algd^{-k}$.  Then $N$ has a subset $R$ with width $r$.
\end{lemma}

\begin{proof}
  We use induction on the cardinality of the set.  The base case
  $|N|=1$ is trivial since then $x=0$.  Assume the lemma holds for all
  $|N| < n$, and suppose $|\tilde{N}|=n$.  Let $\rho^* = \min_{j \in
  \tilde{N}} \rho_j$ and $j^* = \argmin_{j \in \tilde{N}} \rho_j$.  We
  can see $\rho^*$ as the smallest contribution to the width of $\tilde{N}$ and
  $r$ as the portion of the binary expansion of the width of $\tilde{N}$ to the
  right of $\algd^{-k}$.  Then clearly $r$ must be an integer multiple of
  $\rho^*$.  If $r=\rho^*$, $R =\{j^*\}$ is a solution.  Otherwise let
  $N'=\tilde{N} \backslash \{j^*\}$ (so $|N'|=n-1$) and let $R'$ be
  the subset obtained from solving the lemma for set $N'$ of width $r
  - \rho^*$.  Then $R=R'\cup \{j^*\}$.
\end{proof}

We are now able to prove the main theorem:

\begin{theorem}
\label{cceqll}
Any $N$ that is a solution of the Coin Collector's problem for $t =
\rho(N) = n-1$ has a corresponding $\boldlupN$ such that $N =
\nodeset(\boldlupN)$ and $\mu(N) = \min_{\smalll} \Cost(\p, \boldl, f)
- \CostConst(\p,f)$.
\end{theorem}

\begin{proof}
  By monotonicity of the penalty function, any optimal solution
  satisfies the Kraft inequality with equality.  Thus all optimal
  length distribution nodesets have $\rho(\nodeset(\boldl)) = n-1$.
  Suppose $N$ is a solution to the Coin Collector's problem but is not
  a valid nodeset of a length distribution.  Then there exists an
  $(i,l)$ with $l>1$ such that $(i,l) \in N$ and $(i,l-1) \in I
  \backslash N$.  Let $R' = N \cup \{(i,l-1)\} \backslash \{(i,l)\}$.
  Then $\rho(R')=n-1+\algd^{-l}$ and, due to convexity, $\mu(R') \leq
  \mu(N)$.  Thus, using Lemma~\ref{lllemma} with $k=0$,
  $x=n-1$, and $r=\algd^{-l}$, there exists an $R \subset R'$ such
  that $\rho(R)=\algd^{-l}$ and $\mu(R' \backslash R) < \mu(R') \leq
  \mu(N)$.  Since we assumed $N$ to be an optimal solution
  of the Coin Collector's problem, this is a contradiction, and thus
  any optimal solution of the Coin Collector's problem corresponds to
  an optimal length distribution.
\end{proof}

Note that the generality of this algorithm makes it trivially
extensible to problems of the form $\sum_i f_i(l_i,p_i)$ for $n$
different functions $f_i$.  This might be applicable if we desire a
nonlinear weighting for codewords --- such as an additional utility
weight --- in addition to and possibly independent of codeword length
and probability.  

Because the Coin Collector's problem is linear in time and space, the
overall algorithm finds an optimal code in $\order(nl_{\max})$ time
and space for any ``well-behaved'' $f(l_i,p_i)$, that is, any $f$ of
the form specified for which same-width inputs would automatically be
presorted by weight for the Coin Collector's problem.

The complexity of the algorithm in terms of $n$ alone depends on the
structure of both $f$ and~$\p$, because, if we can upper-bound the
maximum length codeword, we can run the Package-Merge algorithm with
fewer input nodes.  In addition, if $f$ is not ``well-behaved,'' input
to the Package-Merge algorithm might need to be sorted.

To quantify these behaviors, we introduce one definition and recall
another:

\begin{definition} A (coding) problem space is called a {\defn flat
class} if there exists a constant upper bound $u$ such that
$\frac{\max_i l_i}{\log n} < u$ for any solution $\boldl$.
\end{definition}

\noindent 
For example, the space of linear
Huffman coding problems with all $p_i \geq \frac{1}{2n}$ is a flat class.
(This may be shown using \cite{Buro}.)

Recall Definition~\ref{difmon} given in Section~\ref{intro}: A cost
function $f(l,p)$ and its associated penalty $\Cost$ are {\defn
differentially monotonic} or {\defn d.m.} if, for every $l>1$,
whenever $f(l-1,p_i)$ is finite and $p_i > p_j$,
$f(l,p_i)-f(l-1,p_i) > f(l,p_j)-f(l-1,p_j)$.
This implies that $f$ is continuous in~$\p$ at all but a countable
number of points.  Without loss of generality, we consider only cases
in which it is continuous everywhere.

If $f(l,p)$ is differentially monotonic, then there is no need to sort
the input nodes for the algorithm.  Otherwise, sorting occurs on
$l_{\max}$ rows with $\order(n \log n)$ on each row, $\order(n
l_{\max} \log n)$ total.  Also, if the problem space is a flat class,
$l_{\max}$ is $\order(\log n)$; it is $\order(n)$ in general.  Thus
time complexity for this solution ranges from $\order(n \log n)$ to
$\order(n^2 \log n)$ with space requirement $\order(n \log n)$ to
$\order(n^2)$; see Table~\ref{complexity} for details.  As indicated
in the table, space complexity can be reduced in differentially
monotonic instances.

\begin{table}
$$
\begin{array}{l|cc}
\mbox{problem type}         & \mbox{time}       & \mbox{space} \\ \hline
\mbox{flat, d.m.}         & \order(n \log n)  & \order(n \log n) \\
\qquad \mbox{space-optimized}                 & \order(n \log n)   & \order(n) \\
\hline
\mbox{not flat, d.m.}     & \order(n^2)       & \order(n^2) \\
\qquad \mbox{space-optimized}                 & \order(n^2)        & \order(n)\\
\hline
\mbox{flat, not d.m.}     & \order(n \log^2 n) & \order(n \log n)\\
\hline
\mbox{not flat, not d.m.} & \order(n^2 \log n) & \order(n^2) \\
\end{array}$$
\caption{Complexity for various types of inputs 
{($\mbox{d.m. = differentially monotonic}$)}}
\label{complexity}
\end{table}

\subsection{A Linear-Space Algorithm}
\label{linear}

Note that the length distribution returned by the algorithm need not
have the property that $l_i \leq l_j$ whenever $i < j$.  For example,
if $p_i = p_j$, we are guaranteed no particular inequality relation
between $l_i$ and $l_j$ since we did not specify a method for breaking
ties.  Also, even if all $p_i$ were distinct, there are cost functions
for which we would expect the inequality relation reversed from the
linear case.  An example of this is $f(l_i, p_i) = p_i^{-1} 2^{l_i}$,
although this represents no practical problem that the author is aware
of.  

Practical cost functions will, given a probability distribution for
nonincreasing $p_i$, generally have at least one optimal code of
monotonically nondecreasing length.  Differentially monotonicity is a
sufficient condition for this, and we can improve upon the algorithm
by insisting that the problem be differentially monotonic and all
entries $p_i$ in $\p$ be distinct; the latter condition we later
relax.  The resulting algorithm uses only linear space and quadratic
time.  First we need a definition:

\begin{definition} A {\defn monotonic} nodeset, $N$, is one with the following properties:
\begin{eqnarray}
(i,l) \in N \Rightarrow (i+1,l) \in N& \for i<n \label{firstprop} \\
(i,l) \in N \Rightarrow (i,l-1) \in N& \for l>1 \label{validlen} 
\end{eqnarray}
This definition is equivalent to that given in \cite{LaHi}.
\end{definition}

An example of a monotonic nodeset is the set of nodes
enclosed by the dashed line in Fig.~\ref{ABCD}.  Note that a nodeset
is monotonic only if it corresponds to a length distribution
$\boldl$ with lengths sorted in nondecreasing order.

\begin{lemma}
\label{dmlemma}
If a problem is differentially monotonic and monotonically
increasing and convex in each $l_i$, and if $\p$ has no repeated values, then
any optimal solution $N=\os(I,n-1)$ is monotonic.
\end{lemma}

\begin{proof}
The second monotonic property (\ref{validlen}) was proved for optimal
nodesets in Theorem~\ref{cceqll}, and the first is now proved with a
simple exchange argument, as in \cite[pp.~97--98]{CoTh}.  Suppose we
have optimal $N$ that violates the first property (\ref{firstprop}).
Then there exist unequal $i$ and $j$ such that $p_i < p_j$ and $l_i <
l_j$ for optimal codeword lengths $\boldl$ ($N = \nodeset(\boldl)$).
Consider $\boldl'$ with lengths for symbols $i$ and $j$ interchanged.
Then
$$
\begin{array}{l}
\Cost(\p, \boldl',f) - \Cost(\p, \boldl,f)  \qquad \\
\quad = \sum_k f(l'_k, p_k) - \sum_k f(l_k,p_k) \\
\quad = \left(f(l_j,p_j) - f(l_i,p_j)\right) - \left(f(l_j,p_i) - f(l_i,p_i)\right) \\
\quad = \sum_{l=l_i+1}^{l_j} \left(M_f(l,p_j)-M_f(l,p_i)\right)\\
\quad < 0
\end{array}
$$
where we recall that $M_f(l,p)\das f(l,p)-f(l-1,p)$
and the final inequality is due to differential monotonicity.
However, this implies that $\boldl$ is not an optimal code,
and thus we cannot have an optimal nodeset without monotonicity unless
values in $\p$ are repeated.
\end{proof}

Taking advantage of this relation to trade off a constant factor of
time for drastically reduced space complexity has been done in
\cite{Larm} for the case of the length-limited (linear) penalty (\ref{llhuff}).  We now extend this to all convex
differentially monotonic cases.

Note that the total width of items that are each less than or equal to
width $\rho$ is less than $2n\rho$.  Thus, when we are processing
items and packages of width $\rho$, fewer than $2n$ packages are
kept in memory.  The key idea in reducing space complexity is to keep
only four attributes of each package in memory instead of the full
contents.  In this manner, we use linear space while retaining enough
information to reconstruct the optimal nodeset in algorithmic
postprocessing.

Define $l_\mid \das \lfloor \frac{1}{2} (l_{\max}+1) \rfloor$.
Package attributes allow us to divide the problem into two subproblems
with total complexity that is at most half that of the original
problem. For each package $S$, we retain the following attributes:
\begin{enumerate}
\item Weight: $\mu(S) \das \sum_{(i,l) \in S} \mu(i,l)$
\item Width: $\rho(S) \das \sum_{(i,l) \in S} \rho(i,l)$
\item Midct: $\nu(S) \das |S \cap I_{\mid}|$
\item Hiwidth: $\psi(S) \das \sum_{(i,l) \in S \cap I_{\hit}} \rho(i,l)$
\end{enumerate}
where $I_{\hi} \das \{ (i,l)~|~l>l_{\mid} \}$ and $I_{\mid}
\das \{ (i,l)~|~l=l_\mid \}$.  We also define $I_\lo \das
\{(i,l)~|~l<l_\mid \}$.

This retains enough information to complete the ``first run'' of the
algorithm with $\order(n)$ space.  The result
will be the package attributes for the optimal nodeset $N$.  Thus, at the
end of this first run, we know the value for $m = \nu(N)$, and we can
consider $N$ as the disjoint union of four sets, shown in
Fig.~\ref{ABCD}:
\begin{figure*}
\begin{center}
\resizebox{13cm}{!}{\includegraphics{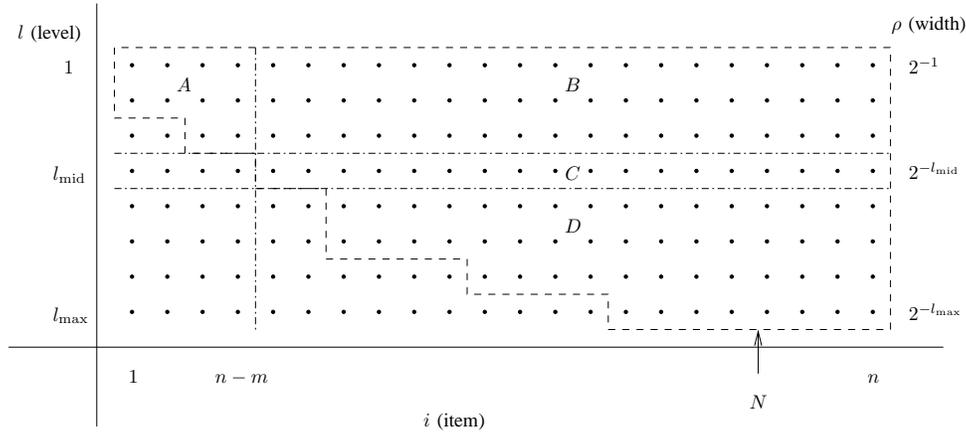}}
\caption{The set of nodes $I$, an optimal nodeset $N$, and disjoint subsets $A$, $B$, $C$, $D$}
\label{ABCD}
\end{center}
\end{figure*}
\begin{enumerate}
\item $A$ = nodes in $N \cap I_\lo$ with indices in $[1,n-m]$,
\item $B$ = nodes in $N \cap I_\lo$ with indices in $[n-m+1,n]$,
\item $C$ = nodes in $N \cap I_\mid$,
\item $D$ = nodes in $N \cap I_\hi$.
\end{enumerate}
Due to monotonicity of $N$, it is trivial that $C = [n-m+1, n] \times
\{l_\mid\}$ and $B = [n-m+1,n] \times [1, l_\mid-1]$.  Note then that
$\rho(C) = m\algd^{-l_\midt}$ and $\rho(B) = m[1-\algd^{-(l_{\midt}-1)}]$.
Thus we need merely to recompute which nodes are in $A$ and in $D$.

Because $D$ is a subset of $I_{\hi}$, $\rho(D) = \psi(N)$ and $\rho(A)
= \rho(N) - \rho(B) - \rho(C) - \rho(D)$.  Given their respective
widths, $A$ is a minimal weight subset of $[1,n-m] \times
[1,l_{\mid}-1]$ and $D$ is a minimal weight subset of $[n-m+1, n]
\times [l_{\mid}+1,l_{\max}]$.  The nodes at each level of $A$ and
$D$ may be found by recursive calls to the algorithm.  In doing so, we
use only $\order(n)$ space.  Time complexity, however, remains the
same; we replace one run of an algorithm on $nl_{\max}$ nodes with a
series of runs, first one on $nl_{\max}$ nodes, then two on an average
of at most $\frac{1}{4}nl_{\max}$ nodes each, then four on
$\frac{1}{16}nl_{\max}$, and so forth.  Formalizing this analysis:

\begin{theorem}
The above recursive algorithm for generalized quasiarithmetic convex coding has
$\order(nl_{\max})$ time complexity. \cite{LaHi}
\end{theorem}

\begin{proof}
As indicated, this recurrence relation is considered and proved in
\cite[pp.~472--473]{LaHi}, but we analyze it here for completeness.
To find the time complexity, set up the following recurrence relation:
Let $T(n,l)$ be the worst case time to find the minimal weight subset
of $[1,n] \times [1,l]$ (of a given width), assuming the subset is
monotonic.  Then there exist constants $c_1$ and $c_2$ such that, if
we define $\hat{l} \das l_{\mid} - 1 \leq \lfloor \frac{l}{2}
\rfloor$ and $\check{l} \das l - \hat{l} - 1 \leq \lfloor
\frac{l}{2} \rfloor$, and we let an adversary choose the corresponding
$\hat{n} + \check{n} = n$,
$$
\begin{array}{rcll}
T(n,l) &\leq& c_1 n &\mbox{for } l < 3 \\
T(n,l) &\leq& c_2 n l + T(\hat{n}, \hat{l}) + T(\check{n}, \check{l})
&\mbox{for } l \geq 3 ,
\end{array}
$$
where $l<3$ is the base case.  Then $T(n,l) = \order(\tau(n,l))$,
where $\tau$ is any function satisfying the recurrence
$$
\begin{array}{rcll}
\tau(n,l) &\geq& c_1 n &\mbox{for } l < 3  \\
\tau(n,l) &\geq& c_2 n l + \tau(\hat{n}, \frac{l}{2}) + 
\tau(n-\hat{n}, \frac{l}{2}) &\mbox{for } l \geq 3,
\end{array}
$$
which $\tau(n,l) = (c_1 + 2c_2)nl$ does.  Thus, the time
complexity is $\order(nl_{\max})$.
\end{proof}

The overall complexity is $\order(n)$ space and $\order(nl_{\max})$ time
--- $\order(n \log n)$ considering only flat classes, $\order(n^2)$ in
general, as in Table~\ref{complexity}.

However, the assumption of distinct $p_i\s$ puts an undesirable
restriction on our input.  In their original algorithm from
\cite{LaHi}, Larmore and Hirschberg suggest modifying the
probabilities slightly to make them distinct, but this is
unnecessarily inelegant, as the resulting algorithm has the drawbacks
of possibly being slightly nonoptimal and being nondeterministic; that
is, different implementations of the algorithm could result in the
same input yielding different outputs.  A deterministic variant of
this approach could involve modifications by multiples of a suitably
small variable $\epsilon > 0$ to make identical values distinct.  In
\cite{LaPr2}, another method of tie-breaking is presented for
alphabetic length-limited codes.  Here, we present a simpler
alternative analogous to this approach, one which is both deterministic
and applicable to all differentially monotonic instances.

Recall that $\p$ is a nonincreasing vector.  Thus items of a given
width are sorted for use in the Package-Merge algorithm; use this
order for ties.  For example, if we use the nodes in
Fig.~\ref{nodesetnum} --- $n=4$, $f(l,p)=pl^2$ --- with probability
$\p = (0.5, 0.2, 0.2, 0.1)$, then nodes $(4,3)$ and $(3,3)$ are the first
to be paired, the tie between $(2,3)$ and $(3,3)$ broken by order.
Thus, at any step, all identical-width items in one package have
adjacent indices.  Recall that packages of items will be either in the
final nodeset or absent from it as a whole.  This scheme then prevents
any of the nonmonotonicity that identical $p_i\s$ might bring about.

In order to ensure that the algorithm is fully deterministic ---
whether or not the linear-space version is used --- the manner in
which packages and single items are merged must also be taken into
account.  We choose to merge nonmerged items before merged items in
the case of ties, in a similar manner to the two-queue bottom-merge
method of Huffman coding\cite{Schw,Leeu}.  Thus, in our example, the
node $(1,2)$ is chosen whereas the package of items $(4,3)$ and
$(3,3)$ is not.  This leads to the optimal length vector $\boldl = (2, 2,
2, 2)$, rather than $\boldl = (1, 2, 3, 3)$ or $\boldl = (1, 3, 2, 3)$,
which are also optimal.  As in bottom-merge Huffman coding, the code
with the minimum reverse lexicographical order among optimal codes is
the one produced.  This is also the case if we use
the position of the ``last'' node in a package (in terms of the
value of $nl+i$) in order to choose those with lower values, as in
\cite{LaPr2}.  However, the above approach, which is easily shown to be
equivalent via induction, eliminates the need for keeping track
of the maximum value of $nl+i$ for each package.

\subsection{Further Refinements}
\label{refine}

In this case using a bottom-merge-like coding method has an additional
benefit: We no longer need assume that all $p_i \neq 0$ to assure that
the nodeset is a valid code.  In finding optimal binary codes, of
course, it is best to ignore an item with $p_i = 0$.  However,
consider nonbinary output alphabets, that is, $\infd > 2$.  As in
Huffman coding for such alphabets, we must add ``dummy'' values of
$p_i = 0$ to assure that the optimal code has the Kraft inequality
satisfied with equality, an assumption underlying both the Huffman
algorithm and ours.  The number of dummy values needed is
$\mod(\infd-n,\infd-1)$ where $\mod(x,y) \das x - y \lfloor
\frac{x}{y} \rfloor$ and where the dummy values each consist of
$l_{\max}$ nodes, each node with the proper width and with weight $0$.
With this preprocessing step, finding an optimal code should proceed
similarly to the binary case, with adjustments made for both the
Package-Merge algorithm and the overall coding algorithm due to the
formulation of the Kraft inequality and maximum length.  A complete
algorithm is available, with proof of correctness, in \cite{Baer20}.

Note that we have assumed for all variations of this algorithm that we
knew a maximum bound for length, although in the overall complexity
analysis for binary coding we assumed this was $n-1$ (except for flat
classes).  We now explore a method for finding better upper bounds and
thus a more efficient algorithm.  First we present a definition due to
Larmore:

\begin{definition} 
Consider penalty functions $f$ and $g$.  We say that $g$ is {\defn
flatter} than $f$ if, for probabilities $p$ and $p'$ and positive
integers $l$ and $l'$ where $l' > l$, $M_g(l,p) M_f(l',p') \leq
M_f(l,p) M_g(l',p')$ (where, again, $M_f(l,p) \das f(l,p)-f(l-1,p)$)
\cite{Larm}.
\end{definition} 

A consequence of the Convex Hull Theorem of \cite{Larm} is that, given
$g$ flatter than $f$, for any $\p$, there exist $f$-optimal $\boldlf$
and $g$-optimal $\boldlg$ such that $\boldlf$ is greater
lexicographically than $\boldlg$ (again, with lengths sorted largest
to smallest).  This explains why the word ``flatter'' is used.

Thus, for penalties flatter than the linear penalty, we can obtain a
useful upper bound, reducing complexity.  All convex quasiarithmetic
penalties are flatter than the linear penalty.  (There are some
generalized quasiarithmetic convex coding penalties that are not
flatter than the linear penalty --- e.g., $f(l_i,p_i) = l_i p_i^2$ ---
and some flatter penalties that are not Campbell/quasiarithmetic ---
e.g., $f(l_i,p_i) = 2^{l_i} (p_i + 0.1 \sin \pi p_i)$ --- so no other
similarly straightforward relation exists.)  For most penalties we
have considered, then, we can use the upper bounds in \cite{Buro} or
the results of a pre-algorithmic Huffman coding of the symbols to
find an upper bound on codeword length.

A problem in which pre-algorithmic Huffman coding would be useful is
delay coding, in which the quadratic penalty (\ref{quadratic}) is
solved for $\order(n^2)$ values of $\alpha$ and $\beta$\cite{Larm}. In
this application, only one traditional Huffman coding would be
necessary to find an upper bound for all quadratic cases.

With other problems, we might wish to instead use a mathematically
derived upper bound.  Using the maximum unary codeword length of $n-1$
and techniques involving the Golden Mean, $\Phi \das
\frac{\sqrt{5}+1}{2}$, Buro in \cite{Buro} gives the upper limit of
length for a (standard) binary Huffman codeword as
$$\min\left\{\left\lfloor \log_\Phi \left (\frac{\Phi+1}{p_{n} \Phi +
        p_{n-1} }\right ) \right \rfloor, n-1 \right \} $$
which would thus be an upper limit on codeword length for
the minimal optimal code obtained using any flatter penalty function,
such as a convex quasiarithmetic function.  This may be used to
reduce complexity, especially in a case in which we encounter a flat
class of problem inputs.  

In addition to this, one can improve this algorithm by
adapting the binary length-limited Huffman coding techniques of Moffat
(with others) in \cite{KMT,LiMo,MTK,TuMo,TuMo2}.  We do not explore
these, however, as these cannot improve asymptotic results with the
exception of a few special cases.  Other approaches to length-limited
Huffman coding with improved algorithmic complexity\cite{AST,Schi} are
not suited for extension to nonlinear penalties.

\section{Conclusion}
\label{conclusion}

With a similar approach to that taken by Shannon for Shannon entropy
and Campbell for R\'{e}nyi entropy, one can show redundancy bounds and
related properties for optimal codes using Campbell's quasiarithmetic
penalties and generalized entropies.  For convex quasiarithmetic costs,
building upon and refining Larmore and Hirschberg's methods, one can construct efficient algorithms for finding an optimal code.  Such algorithms can
be readily extended to the generalized quasiarithmetic convex class of
penalties, as well as to the delay penalty, the latter of which
results in more quickly finding an optimal code for delay channels.

One might ask whether the aforementioned properties can be extended;
for example, can improved redundancy bounds similar to \cite{BaSi, BlMc,
CaDe, Tane} be found?  It is an intriguing question, albeit one that seems
rather difficult to answer given that such general penalties lack a
Huffman coding tree structure.  In addition, although we know that
optimal codes for infinite alphabets exist given the aforementioned
conditions, we do not know how to find them.  This, as with many
infinite alphabet coding problems, remains open.

It would also be interesting if the algorithms could be extended to
other penalties, especially since complex models of queueing can lead
to other penalties aside from the delay penalty mentioned here.  Also,
note that the monotonicity property of the examples we consider
implies that the resulting optimal code can be alphabetic, that is,
lexicographically ordered by item number.  If we desire items to be in
a lexicographical order different from that of probability, however,
the alphabetic and nonalphabetic cases can have different solutions.
This was discussed for the length-limited penalty in \cite{LaPr2}; it
might be of interest to generalize it to other penalties using similar
techniques and to prove properties of alphabetic codes for such
penalties.

\section*{Acknowledgments}

The author wishes to thank Thomas Cover, John Gill, and Andrew
Brzezinski for feedback on this paper and research, as well as the
anonymous reviewers for their numerous suggestions for improvement.
Discussions and comments from the Stanford Information Theory Group
and Benjamin Van Roy are also greatly appreciated, as is encouragement
from Brian Marcus.

\appendix[The Package-Merge Algorithm]

Here we illustrate and prove the correctness of a recursive version of
Package-Merge algorithm for solving the Coin Collector's problem.
This algorithm was first presented in \cite{LaHi}, which also has a
linear-time iterative implementation.

Restating the Coin Collector's problem:
\begin{equation}
\begin{array}{ll}
\mbox{Minimize} \,_{\{B \subseteq \{1,\ldots,m\}\}} & \sum_{i \in B} \mu_i  \\
\mbox{subject to } & \sum_{i \in B} \rho_i = t \\
\mbox{where } & m \in \Z_+ \\ 
& \mu_i \in \R \\
& \rho_i \in \algd^\Z \\
& t \in \R_+
\end{array} 
\end{equation}
In our notation, we use $i \in \{1,\ldots,m\}$ to denote both the
index of a coin and the coin itself, and $\I$ to represent the $m$
items along with their weights $\{\mu_i\}$ and widths $\{\rho_i\}$.
The optimal solution, a function of total width $t$ and items $\I$, is
denoted $\os(\I,t)$.

Note that we assume the solution exists but might not be unique.  In
the case of distinct solutions, tie resolution for minimizing
arguments may for now be arbitrary or rule-based; we clarify this in
Subsection~\ref{linear}.  A modified version of the algorithm
considers the case where a solution might not exist, but this is not
needed here.  Because a solution exists, assuming $t>0$, $t=\omega 
t_\power$ for some unique odd $\omega \in \Z$ and $t_\power \in \algd^\Z$.  (Note
that $t_\power$ need not be an integer.  If $t = 0$, $\omega$ and $t_\power$ are not
defined.)

\noindent {\bf Algorithm variables} \\
At any point in the algorithm, given nontrivial $\I$ and $t$, we use the following definitions: 

\begin{tabular}{rcl}
Remainder & & \\ 
$t_\power$ & $\definedas$ & the unique $x \in \algd^\Z$ such \\
& & that $\frac{t}{x}$ is
an odd integer \\[2pt]
Minimum width & & \\
$\rho^*$ & $\definedas$ & $\min_{i \in \I} \rho_i$ (note
$\rho^* \in \algd^\Z$) \\[2pt]
Small width set & & \\
$\I^*$ & $\definedas$ & $\{i~|~\rho_i = \rho^*\}$ \\
& & (by definition, $|\I^*| \geq 1$) \\[2pt]
``First'' item & & \\
$i^*$ & $\definedas$ & $\argmin_{i \in \I^*} \mu_i$ \\[2pt]
``Second'' item & & \\
$i^{**}$ & $\definedas$ & $\argmin_{i \in \I^* \backslash \{i^*\}} \mu_i$ \\
& & (or null $\Lambda$ if $|\I^*| = 1$) 
\end{tabular}

${}$

\noindent Then the following is a recursive description
of the algorithm:

\noindent {\bf Recursive Package-Merge Procedure} \cite{LaHi}

{\it Basis.  $t = 0$:}  $\os (\I,t)$ is the empty set.

{\it Case 1.  $\rho^* = t_\power$ and $\I \neq \emptyset$:}  $\os(\I,t) =
\os (\I \backslash \{i^*\},t-\rho^*) \cup \{i^*\}$.

{\it Case 2a.  $\rho^* < t_\power$, $\I \neq \emptyset$, and $|\I^*| = 1$:}
$\os(\I,t) = \os(\I \backslash \{i^*\}, t)$.

{\it Case 2b.  $\rho^* < t_\power$, $\I \neq \emptyset$, and $|\I^*| > 1$:} 
Create $i'$, a new item with weight $\mu_{i'} = \mu_{i^*} + \mu_{i^{**}}$ and
width $\rho_{i'} = \rho_{i^*} +\rho_{i^{**}} = 2\rho^*$.  This new item is
thus a combined item, or {\defn package}, formed by combining items $i^*$
and $i^{**}$.  Let $S' = \os(\I \backslash \{i^*,i^{**}\} \cup \{i'\}, t)$
(the optimization of the packaged version).  If $i' \in S'$, then
$\os(\I,t) = S' \backslash \{i'\} \cup \{i^*,i^{**}\}$; otherwise,
$\os(\I,t) = S'$.

${}$

\begin{figure*}[t]
\begin{center}
\resizebox{12cm}{!}{\includegraphics{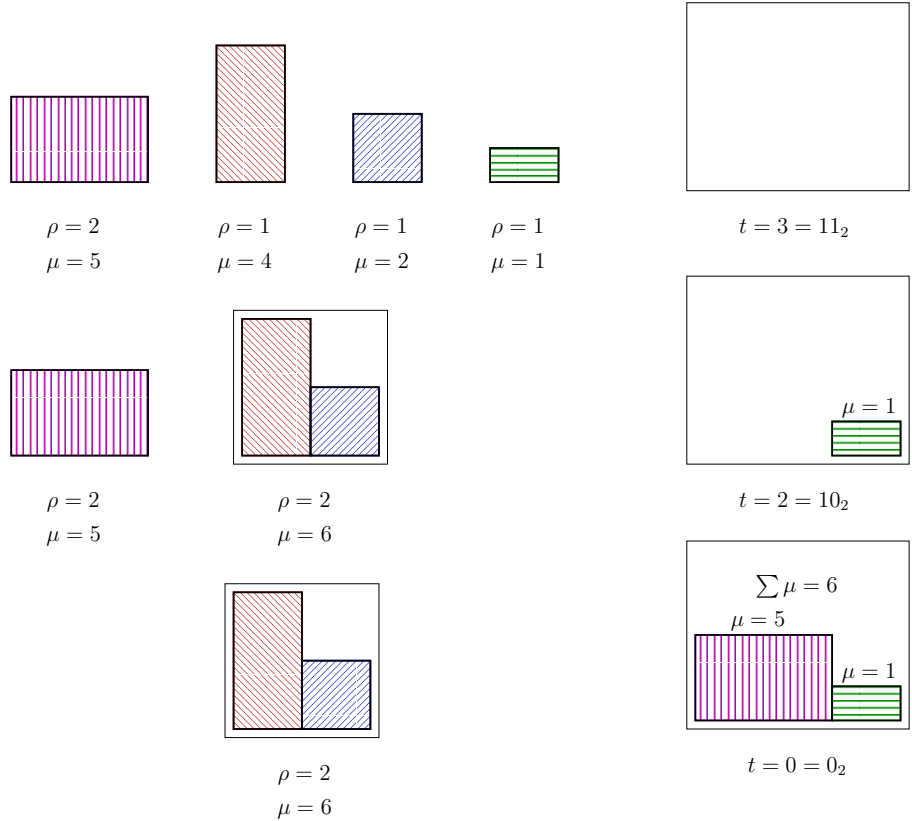}}
\caption{A simple example of the Package-Merge algorithm}
\label{pm}
\end{center}
\end{figure*}

\begin{theorem}
If an optimal solution to the Coin Collector's problem exists, the
above recursive (Package-Merge) algorithm will terminate with an
optimal solution. 
\end{theorem}

\begin{proof}
We show that the Package-Merge algorithm produces
an optimal solution via induction on the depth of the recursion.  The
basis is trivially correct, and each inductive case reduces the number of
items by one.  The inductive hypothesis on $t \geq 0$ and $\I \neq
\emptyset$ is that the algorithm is correct for any problem instance that
requires fewer recursive calls than instance $(\I,t)$.

If $\I=\emptyset$ and $t \neq 0$, or if $\rho^* > t_\power > 0$, then there
is no solution to the problem, contrary to our assumption.  Thus all
feasible cases are covered by those given in the procedure.  Case 1
indicates that the solution must contain an odd number of elements of
width~$\rho^*$.  These must include the minimum weight item in $\I^*$,
since otherwise we could substitute one of the items with this
``first'' item and achieve improvement.  Case~2 indicates that the
solution must contain an even number of elements of width~$\rho^*$.
If this number is $0$, neither $i^*$ nor $i^{**}$ is in the solution.
If it is not, then they both are.  If $i^{**} = \Lambda$, the number
is $0$, and we have Case 2a.  If not, we may ``package'' the items,
considering the replaced package as one item, as in Case 2b.  Thus the
inductive hypothesis holds and the algorithm is correct.
\end{proof}

Fig.~\ref{pm} presents a simple example of this algorithm at work,
finding minimum total weight items of total width $t=3$ (or, in
binary, $11_2$).  In the figure, item width represents numeric width
and item area represents numeric weight.  Initially, as shown in the
top row, the minimum weight item with width $\rho_{i^*} = t_\power = 1$ is
put into the solution set.  Then, the remaining minimum width items
are packaged into a merged item of width $2$ ($10_2$).  Finally, the
minimum weight item/package with width $\rho_{i^*} = t_\power = 2$ is added
to complete the solution set, which is now of weight $6$.  The
remaining packaged item is left out in this case; when the algorithm
is used for coding, several items are usually left out of the optimal set.


\begin{thebibliography}{10}
\providecommand{\url}[1]{#1}
\csname url@rmstyle\endcsname
\providecommand{\newblock}{\relax}
\providecommand{\bibinfo}[2]{#2}
\providecommand\BIBentrySTDinterwordspacing{\spaceskip=0pt\relax}
\providecommand\BIBentryALTinterwordstretchfactor{4}
\providecommand\BIBentryALTinterwordspacing{\spaceskip=\fontdimen2\font plus
\BIBentryALTinterwordstretchfactor\fontdimen3\font minus
  \fontdimen4\font\relax}
\providecommand\BIBforeignlanguage[2]{{%
\expandafter\ifx\csname l@#1\endcsname\relax
\typeout{** WARNING: IEEEtran.bst: No hyphenation pattern has been}%
\typeout{** loaded for the language `#1'. Using the pattern for}%
\typeout{** the default language instead.}%
\else
\language=\csname l@#1\endcsname
\fi
#2}}

\bibitem{Huff}
D.~Huffman, ``A method for the construction of minimum-redundancy codes,''
  \emph{Proc. IRE}, vol.~40, no.~9, pp. 1098--1101, Sept. 1952.

\bibitem{Abr01}
J.~Abrahams, ``Code and parse trees for lossless source encoding,''
  \emph{Communications in Information and Systems}, vol.~1, no.~2, pp.
  113--146, Apr. 2001.

\bibitem{Humb0}
P.~Humblet, ``Source coding for communication concentrators,'' Ph.D.
  dissertation, Massachusetts Institute of Technology, 1978.

\bibitem{Camp}
L.~Campbell, ``Definition of entropy by means of a coding problem,'' \emph{Z.
  Wahrscheinlichkeitstheorie und verwandte Gebiete}, vol.~6, pp. 113--118,
  1966.

\bibitem{McMi}
B.~McMillan, ``Two inequalities implied by unique decipherability,'' \emph{IRE
  Trans. Inf. Theory}, vol. IT-2, no.~4, pp. 115--116, Dec. 1956.

\bibitem{Larm}
L.~Larmore, ``Minimum delay codes,'' \emph{SIAM J. Comput.}, vol.~18, no.~1,
  pp. 82--94, Feb. 1989.

\bibitem{Humb2}
P.~Humblet, ``Generalization of {Huffman} coding to minimize the probability of
  buffer overflow,'' \emph{IEEE Trans. Inf. Theory}, vol. IT-27, no.~2, pp.
  230--232, Mar. 1981.

\bibitem{Kapu}
J.~Kapur, ``Noiseless coding theorems for different measures of entropy,''
  \emph{J. Comb., Inf. Syst. Sci.}, vol.~13, no. 3--4, pp. 114--126, 1989.

\bibitem{Flor}
C.~Flores, ``Encoding of bursty sources under a delay criterion,'' Ph.D.
  dissertation, University of California, Berkeley, 1983.

\bibitem{HKT}
T.~Hu, D.~Kleitman, and J.~Tamaki, ``Binary trees optimum under various
  criteria,'' \emph{SIAM J. Appl. Math.}, vol.~37, no.~2, pp. 246--256, Apr.
  1979.

\bibitem{Park}
D.~Parker, Jr., ``Conditions for optimality of the {Huffman} algorithm,''
  \emph{SIAM J. Comput.}, vol.~9, no.~3, pp. 470--489, Aug. 1980.

\bibitem{BaerI06}
M.~Baer, ``{R\'{e}nyi} to {R\'{e}nyi} --- source coding under siege,'' in
  \emph{Proceedings of the 2006 IEEE International Symposium on Information
  Theory}, 2006, to appear.

\bibitem{Reny}
A.~R{\'{e}}nyi, \emph{A Diary on Information Theory}.\hskip 1em plus 0.5em
  minus 0.4em\relax New York, NY: John Wiley {\&} Sons Inc., 1987, original
  publication: {\it Napl\`{o} az inform\'{a}ci\'{o}elm\'{e}letr\H{o}l},
  Gondolat, Budapest, Hungary, 1976.

\bibitem{LaHi}
L.~Larmore and D.~Hirschberg, ``A fast algorithm for optimal length-limited
  {Huffman} codes,'' \emph{J. ACM}, vol.~37, no.~2, pp. 464--473, Apr. 1990.

\bibitem{Gare}
M.~Garey, ``Optimal binary search trees with restricted maximal depth,''
  \emph{SIAM J. Comput.}, vol.~3, no.~2, pp. 101--110, June 1974.

\bibitem{Itai}
A.~Itai, ``Optimal alphabetic trees,'' \emph{SIAM J. Comput.}, vol.~5, no.~1,
  pp. 9--18, Mar. 1976.

\bibitem{Ren1}
A.~R{\'{e}}nyi, ``Some fundamental questions of information theory,''
  \emph{Magyar Tudom{\'{a}}nyos Akad{\'{e}}mia III. Osztalyanak
  K{\"{o}}zlemenei}, vol.~10, no.~1, pp. 251--282, 1960.

\bibitem{Acz3}
J.~Acz{\'{e}}l, ``On {Shannon}'s inequality, optimal coding, and
  characterizations of {Shannon}'s and {R\'{e}nyi}'s entropies,'' in
  \emph{Symposia Mathematica}, vol.~15.\hskip 1em plus 0.5em minus 0.4em\relax
  New York, NY: Academic Press, 1973, pp. 153--179.

\bibitem{LTZ}
T.~Linder, V.~Tarokh, and K.~Zeger, ``Existence of optimal prefix codes for
  infinite source alphabets,'' \emph{IEEE Trans. Inf. Theory}, vol. IT-43,
  no.~6, pp. 2026--2028, Nov. 1997.

\bibitem{Schw}
E.~Schwartz, ``An optimum encoding with minimum longest code and total number
  of digits,'' \emph{Inf. Contr.}, vol.~7, no.~1, pp. 37--44, Mar. 1964.

\bibitem{Baer20}
M.~Baer, ``Twenty (or so) questions: {$D$-ary} bounded-length {Huffman}
  coding,'' preprint available from \url{http://arxiv.org/abs/cs.IT/0602085}.

\bibitem{KaNe}
G.~Katona and T.~Nemetz, ``{Huffman} codes and self-information,'' \emph{IEEE
  Trans. Inf. Theory}, vol. IT-22, no.~3, pp. 337--340, May 1976.

\bibitem{Buro}
M.~Buro, ``On the maximum length of {Huffman} codes,'' \emph{Inf. Processing
  Letters}, vol.~45, no.~5, pp. 219--223, Apr. 1993.

\bibitem{AbMc}
Y.~Abu-Mostafa and R.~McEliece, ``Maximal codeword lengths in {Huffman}
  codes,'' \emph{Computers {\&} Mathematics with Applications}, vol.~39,
  no.~11, pp. 129--134, Oct. 2000.

\bibitem{GaJo}
M.~Garey and D.~Johnson, \emph{Computers and Intractability}.\hskip 1em plus
  0.5em minus 0.4em\relax San Francisco, CA: W.H.~Freeman and Company, 1979.

\bibitem{Manb}
U.~Manber, \emph{Introduction to Algorithms}.\hskip 1em plus 0.5em minus
  0.4em\relax Reading, MA: Addison-Wesley, 1989.

\bibitem{CoTh}
T.~Cover and J.~Thomas, \emph{Elements of Information Theory}.\hskip 1em plus
  0.5em minus 0.4em\relax New York, NY: Wiley-Interscience, 1991.

\bibitem{LaPr2}
L.~Larmore and T.~Przytycka, ``A fast algorithm for optimal height-limited
  alphabetic binary-trees,'' \emph{SIAM J. Comput.}, vol.~23, no.~6, pp.
  1283--1312, Dec. 1994.

\bibitem{Leeu}
J.~{van Leeuwen}, ``On the construction of {Huffman} trees,'' in \emph{Proc.
  3rd Int. Colloquium on Automata, Languages, and Programming}, July 1976, pp.
  382--410.

\bibitem{KMT}
J.~Katajainen, A.~Moffat, and A.~Turpin, ``A fast and space-economical
  algorithm for length-limited coding,'' in \emph{Proceedings of the
  International Symposium on Algorithms and Computation}, Dec. 1995, p. 1221.

\bibitem{LiMo}
M.~Liddell and A.~Moffat, ``Incremental calculation of optimal
  length-restricted codes,'' in \emph{Proceedings, IEEE Data Compression
  Conference}, Apr. 2002, pp. 182--191.

\bibitem{MTK}
A.~Moffat, A.~Turpin, and J.~Katajainen, ``Space-efficient construction of
  optimal prefix codes,'' in \emph{Proceedings, IEEE Data Compression
  Conference}, Mar. 28--30, 1995, pp. 192--202.

\bibitem{TuMo}
A.~Turpin and A.~Moffat, ``Practical length-limited coding for large
  alphabets,'' \emph{The Computer Journal}, vol.~38, no.~5, pp. 339--347, 1995.

\bibitem{TuMo2}
------, ``Efficient implementation of the package-merge paradigm for generating
  length-limited codes,'' in \emph{Proceedings of Computing: The Australasian
  Theory Symposium}, Jan. 29--30, 1996, pp. 187--195.

\bibitem{AST}
A.~Aggerwal, B.~Schieber, and T.~Tokuyama, ``Finding a minimum-weight
  {$k$}-link path on graphs with the concave {Monge} property and
  applications,'' \emph{Discrete and Computational Geometry}, vol.~12, pp.
  263--280, 1994.

\bibitem{Schi}
B.~Schieber, ``Computing a minimum-weight $k$-link path in graphs with the
  concave {Monge} property,'' \emph{Journal of Algorithms}, vol.~29, no.~2, pp.
  204--222, Nov. 1998.

\bibitem{BaSi}
D.~Baron and A.~Singer, ``On the cost of worst case coding length
  constraints,'' \emph{IEEE Trans. Inf. Theory}, vol. IT-47, no.~7, pp.
  3088--3090, Nov. 2001.

\bibitem{BlMc}
A.~Blumer and R.~McEliece, ``The {R\'{e}nyi} redundancy of generalized
  {Huffman} codes,'' \emph{IEEE Trans. Inf. Theory}, vol. IT-34, no.~5, pp.
  1242--1249, Sept. 1988.

\bibitem{CaDe}
R.~Capocelli and A.~{De Santis}, ``On the redundancy of optimal codes with
  limited word length,'' \emph{IEEE Trans. Inf. Theory}, vol. IT-38, no.~2, pp.
  439--445, Mar. 1992.

\bibitem{Tane}
I.~Taneja, ``A short note on the redundancy of degree {$\alpha$},'' \emph{Inf.
  Sci.}, vol.~39, no.~2, pp. 211--216, Sept. 1986.

\end{thebibliography}
\end{document}